\documentclass[twocolumn]{aastex7}
\usepackage[]{units}
\usepackage[update,prepend]{epstopdf}
\usepackage{graphicx}
\usepackage{amssymb}
\usepackage{amsmath}
\usepackage{threeparttable}
\usepackage{hyperref}
\usepackage{color}
\usepackage{gensymb}
\usepackage{mathtools}
\usepackage[T1]{fontenc}

\usepackage{makecell}
\usepackage{bm}

\newcommand{\jj}{J0705$+$0612}

\mathchardef\mhyphen="2D
\submitjournal{ApJ}
\shorttitle{Candidate circumsecondary disk wind}
\shortauthors{Zakamska et al.}
\shortauthors{}

\graphicspath {{figs/} }

\begin{document}

\title{ASASSN-24fw: Candidate gas-rich circumsecondary disk occultation of a main-sequence star}

\author[0000-0001-6100-6869]{Nadia L. Zakamska}
\affiliation{William H.\ Miller III Department of Physics \& Astronomy,
Johns Hopkins University, 3400 N. Charles St., Baltimore, MD 21218, USA}
\affiliation{Institute for Advanced Study, Einstein Dr., Princeton NJ 08540}
\email{zakamska@jhu.edu}

\author[0000-0002-5864-1332]{Gautham Adamane Pallathadka}
\affiliation{William H.\ Miller III Department of Physics \& Astronomy,
Johns Hopkins University, 3400 N. Charles St., Baltimore, MD 21218, USA}
\email{gadaman1@jhu.edu}

\author[0000-0002-3601-133X]{Dmitry Bizyaev}
\affiliation{Apache Point Observatory and New Mexico State University, Sunspot, NM 88349, USA}
\affiliation{Sternberg Astronomical Institute, Moscow State University, Universitetskiy prosp. 13, Moscow, 119234, Russia}
\email{dmbiz@apo.nmsu.edu}

\author[0000-0001-6355-2468]{Jaroslav Merc} 
\affiliation{Astronomical Institute of Charles University, V Hole\v{s}ovi\v{c}k\v{a}ch 2, Prague, 18000, Czech Republic}
\affiliation{Instituto de Astrof\'{\i}sica de Canarias, Calle V\'{\i}a L\'actea, s/n, E-38205 La Laguna, Tenerife, Spain}
\email{jaroslav.merc@gmail.com}

\author[0000-0002-4856-7837]{James E. Owen}
\affiliation{Astrophysics Group, Imperial College London, Prince Consort Road, London SW7 2AZ, UK}
\email{james.owen@imperial.ac.uk}

\author[0000-0001-6533-6179]{Henrique Reggiani}
\affiliation{Gemini Observatory/NSF’s NOIRLab, Casilla 603, La Serena, Chile}
\email{henrique.reggiani@noirlab.edu}

\author[0000-0001-5761-6779]{Kevin C. Schlaufman}
\affiliation{William H.\ Miller III Department of Physics \& Astronomy,
Johns Hopkins University, 3400 N. Charles St., Baltimore, MD 21218, USA}
\email{kschlaufman@jhu.edu}

\author[0000-0003-1034-1557]{Karolina B\k{a}kowska}
\affiliation{Institute of Astronomy, Faculty of Physics, Astronomy and Informatics, Nicolaus Copernicus University in Toru\'{n}, Grudzi\k{a}dzka 5, 87-100 Toru\'{n}, Poland}
\email{bakowska@umk.pl}

\author[0009-0000-1171-3370]{S\l{}awomir Bednarz}
\affiliation{Silesian University of Technology, Akademicka 16, 44-100 Gliwice, Poland}
\email{sbednarz@polsl.pl}

\author[0000-0003-4647-7114]{Krzysztof Bernacki}
\affiliation{Silesian University of Technology, Akademicka 16, 44-100 Gliwice, Poland}
\email{kbernacki@polsl.pl}

\author[0000-0002-9441-0195]{Agnieszka Gurgul}
\affiliation{Institute of Astronomy, Faculty of Physics, Astronomy and Informatics, Nicolaus Copernicus University in Toru\'{n}, Grudzi\k{a}dzka 5, 87-100 Toru\'{n}, Poland}
\email{agn.gurgul@gmail.com}

\author[0000-0002-4176-845X]{Kirsten R. Hall}
\affiliation{Radio and Geoastronomy Division, Center for Astrophysics $\vert$ Harvard \& Smithsonian, Cambridge, MA 02135, USA}
\email{kirsten.hall@cfa.harvard.edu}

\author[0000-0003-0125-8700]{Franz-Josef Hambsch}
\affiliation{Vereniging Voor Sterrenkunde (VVS), Zeeweg 96, B-8200 Brugge, Belgium}
\affiliation{Bundesdeutsche Arbeitsgemeinschaft für Veränderliche Sterne, Munsterdamm 90, D-12169 Berlin, Germany}
\email{hambsch@telenet.be}

\author[0009-0005-1710-6754]{Barbara Joachimczyk}
\affiliation{Institute of Astronomy, Faculty of Physics, Astronomy and Informatics, Nicolaus Copernicus University in Toru\'{n}, Grudzi\k{a}dzka 5, 87-100 Toru\'{n}, Poland}
\email{312255@stud.umk.pl}

\author[0000-0003-4960-7463]{Krzysztof Kotysz}
\affiliation{Astronomical Observatory, University of Warsaw, Al. Ujazdowskie 4, 00-478 Warsaw, Poland}
\affiliation{Astronomical Institute, University of Wroc\l{}aw, ul. Miko\l{}aja Kopernika 11, 51-622 Wroc\l{}aw, Poland}
\email{kotysz@astro.uni.wroc.pl}

\author[0000-0002-1557-0343]{Sebastian Kurowski}
\affiliation{Astronomical Observatory, Jagiellonian University, ul. Orla 171, 30-244 Krak\'{o}w, Poland}
\email{kurowski.sebastian@gmail.com}

\author[0000-0002-0490-1469]{Alexios Liakos}
\affiliation{Institute for Astronomy, Astrophysics, Space Applications and Remote Sensing, National Observatory of Athens, Metaxa \& Pavlou St., GR-15236, Penteli, Athens, Greece}
\email{alliakos@noa.gr}

\author[0000-0001-8916-8050]{Przemys\l{}aw J. Miko{\l}ajczyk}
\affiliation{Astronomical Observatory, University of Warsaw, Al. Ujazdowskie 4, 00-478 Warsaw, Poland}
\affiliation{Astrophysics Division, National Centre for Nuclear Research, Pasteura 7, 02-093 Warsaw, Poland}
\affiliation{Astronomical Institute, University of Wroc\l{}aw, ul. Miko\l{}aja Kopernika 11, 51-622 Wroc\l{}aw, Poland}
\email{mikolajczyk@astro.uni.wroc.pl}

\author[0000-0002-3326-2918]{Erika Pak\v{s}tien\.{e}}
\affiliation{Institute of Theoretical Physics and Astronomy, Vilnius University, Saul\.{e}tekio al. 3, Vilnius, 10257, Lithuania}
\email{erika.pakstiene@tfai.vu.lt}

\author[0000-0002-6495-0676]{Grzegorz Pojma{\'n}ski}
\affiliation{Astronomical Observatory, University of Warsaw, Al. Ujazdowskie 4, 00-478 Warsaw, Poland}
\email{gp@astrouw.edu.pl}

\author[0000-0003-3184-5228]{Adam Popowicz}
\affiliation{Silesian University of Technology, Akademicka 16, 44-100 Gliwice, Poland}
\email{Adam.Popowicz@polsl.pl}

\author[0000-0002-5060-3673]{Daniel E. Reichart}
\affiliation{Department of Physics and Astronomy, University of North Carolina at Chapel Hill, Chapel Hill, NC 27599-3255}
\email{reichart@physics.unc.edu}

\author[0000-0002-9658-6151]{\L{}ukasz Wyrzykowski}
\affiliation{Astronomical Observatory, University of Warsaw, Al. Ujazdowskie 4, 00-478 Warsaw, Poland}
\affiliation{Astrophysics Division, National Centre for Nuclear Research, Pasteura 7, 02-093 Warsaw, Poland}
\email{wyrzykow@gmail.com}

\author[0009-0000-9910-1124]{Justas Zdanavi\v{c}ius}
\affiliation{Institute of Theoretical Physics and Astronomy, Vilnius University, Saul\.{e}tekio al. 3, Vilnius, 10257, Lithuania}
\email{justas.zdanavicius@tfai.vu.lt}

\author[0000-0001-5836-9503]{Micha\l{} \.Zejmo}
\affiliation{Janusz Gil Institute of Astronomy, University of Zielona Gora, Szafrana 2, 65-516 Zielona Gora, Poland}
\email{michalzejmo@gmail.com}

\author[0000-0001-6434-9429]{Pawe{\l} Zieli\'{n}ski}
\affiliation{Institute of Astronomy, Faculty of Physics, Astronomy and Informatics, Nicolaus Copernicus University in Toru\'{n}, Grudzi\k{a}dzka 5, 87-100 Toru\'{n}, Poland}
\email{pzielinski@umk.pl}

\author[0000-0003-3609-382X]{Staszek Zola}
\affiliation{Astronomical Observatory, Jagiellonian University, ul. Orla 171, 30-244 Krak\'{o}w, Poland}
\email{szola@oa.uj.edu.pl}

\begin{abstract}
Dusty disks around planetary and substellar companions in outer reaches of exo-planetary systems can be detected as long-lasting occultations, provided the observer is close to the secondary's orbital plane. Here we report optical spectroscopy with KOSMOS (APO), MagE (Magellan) and GHOST (Gemini-S) of ASASSN-24fw (Gaia 07:05:18.97+06:12:19.4), a 4-magnitude dimming event of a main-sequence star which lasted 8.5 months. We discover multiple low-ionization metal emission lines with velocity dispersion $\lesssim 10$ km s$^{-1}$ blue-shifted by 27 km s$^{-1}$ with respect to the star, as well as kinematically complex Na D absorption. If associated with the occulter, these detections suggest that the occulter is gas-rich. Further, we detect blue-shifted and broad ($\sim 200$ km s$^{-1}$) H$\alpha$ line, which likely originates in the inner circumstellar disk. We confirm the previously reported occultations in 1981 and 1937 seen in historic data, yielding a semi-major axis of the occulter's orbital motion around the star of 14 AU. If the occulter is a circumsecondary disk filling 30-100\% of the Hill radius, we estimate the minimum mass of the secondary to be a few Jupiter masses and a disk mass of 1\% of the mass of the Moon. Given the age of the star ($>2$ Gyr), the disk is unlikely to be a survivor of the planet formation stage and may be a result of a planetary collision. If Na D absorption and/or metal emission lines originate in the disk, the observations presented here are the first discovery of a circumsecondary disk wind or rotation. 
\end{abstract}

\section{Introduction}

Circumstellar disks around young stars dissipate due to photo-evaporation of the material by the radiation of the star on timescales of a few Myr \citep{owen12, hart16}. The planets forming around young stars may be able to retain their circumplanetary disks longer, especially if the planet is in the outer reach of its stellar system. Such young disks would be gas-rich and optically thick in the optical and near-IR \citep{ayli09, ward10, mart11, zhu15, owen16, szul16, mart23}, so with a fortuitous alignment of the orbit and the observer one might expect a long-lasting deep transit. For such gas-rich disk, the relative inclination between the circumplanetary disk and the orbit is not a critical component of observability because the apparent scale height of the gas disk is larger than the star.

Eventually circumplanetary disks accrete onto the planet or dissipate on timescales $\lesssim 10^7$ years \citep{mitc11}, but some fraction can evolve into exoplanetary systems with rings or satellites. The potential observational signatures of such older gas-poor systems have been studied theoretically \citep{barn04, ohta09, schl11, tusn11, zulu15} and in a variety of datasets \citep{brow01, heis15, sant15a}. Most results have been non-detections, but \citet{mama12} discovered an exciting candidate which appears to be razor-thin, perhaps even with ringed structure, producing extraordinarily complex lightcurve as it transited the primary star. For a razor-thin disk to block enough light of the star to be observable in transit, a misalignment between the angular momentum of the exoplanet's orbit and its disk is necessary. \citet{zana16} showed that the disk might be able to maintain this obliquity out to large distances from the exoplanet.

While occultations of young stellar objects by clumps in their circumstellar disks have been known for a decade \citep{ansd16}, main-sequence dipper stars are rare and are only now being identified in modern large Galactic variability surveys \citep{tzan25}, and the origin of the occulting circumstellar material in these objects is not yet understood. Some long-lasting, deep transiting events around relatively normal stars are attributed to disintegrating planets \citep{boya16, nesl17} or planetary collisions \citep{meng14, su22}, with swarms of debris blocking the starlight and heated to produce infrared excess \citep{kenw23, mars23}. The dimming events around main sequence or evolved stars that are best explained by long-lived disks around the much dimmer secondary (planet or low-mass star) are challenging to classify \citep{rapp19} due to the one-off nature of the event, since the transiting body is on a long orbit and there is little possibility of repeat observations. Stand-outs in this category are a periodic event described by \citet{dong14} and \citet{scot14} and another by \citet{ratt15}, which are likely to be circumsecondary disks around low-mass stellar companions to evolved stars. 

ASASSN-24fw (J2016 {\it Gaia} coordinates 07:05:18.97 +06:12:19.4), hereafter \jj, is a deep (4-magnitude) dimming event of a 13th magnitude, 1.4$M_{\odot}$ main-sequence star at 1 kpc from the Sun. The event started in September 2024 \citep{joha24} and concluded in the end of May 2025, as we illustrate below. The event has some of the hallmarks of the long-sought gas-rich circumsecondary disks. First, it is long-lasting, indicating origin in the outer stellar system and a large occulter size. Second, the lightcurve is smooth, in contrast to the rapid variability seen within the \citet{mama12} event, suggesting that the occulter in \jj\ is gas-rich. Finally, it is deep, indicating high optical depth and, again, a large geometric size of the occulter. The star shows an infrared excess in the pre-event WISE data, indicating that there is a circumstellar disk and therefore supporting the possibility of a gas-rich circumsecondary disk. \citet{nair24} analyzed historic Harvard plates and discovered two other events in 1937 and 1981, allowing them to determine the orbital period (44 years) and to predict the duration of the occultations (9 months). 

Here we present follow-up observations of \jj\ and the measurements of the occulter parameters and physical conditions. In Section \ref{sec:pre} we summarize all pre-event data and discuss the star's evolutionary status. In Section \ref{sec:event} we describe our follow-up observations during the occultation. We model the system and discuss our results in Section \ref{sec:model} and we conclude in Section \ref{sec:conclusions}. An independent follow-up and analysis of \jj\ is presented in a simultaneous article by \citet{fore25}, and we briefly comment on their data and results in comparison to ours throughout the paper. Surface gravity in the $\log g$ values is given in cgs units (cm s$^{-2}$). All wavelengths are given in air in the heliocentric frame. All wavelengths of atomic and ionic transitions are from the NIST Atomic Spectra Database \citep{nist24}.

\section{Pre-occultation data}
\label{sec:pre}

\begin{figure}
	\centering
	\includegraphics[width=1\linewidth]{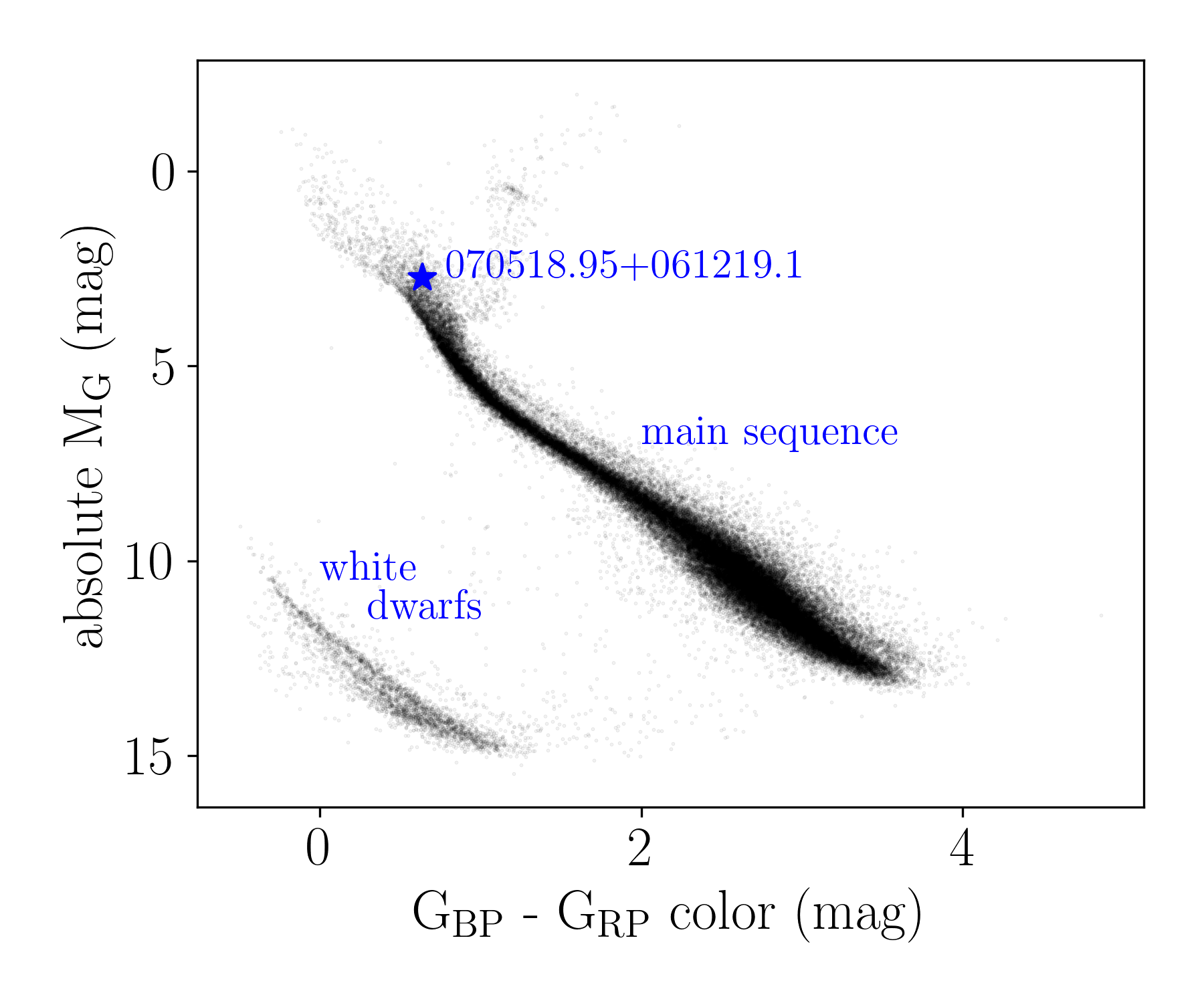}
	\caption{The position of \jj\ on the {\it Gaia} color-magnitude diagram.}
	\label{pic:gaia}
\end{figure}

Here we present the archival observations available for \jj\ prior to the beginning of the occultation event. {\it Gaia} \texttt{source\_id} is 3152916838954800512. Its pre-occultation position on the {\it Gaia} color-magnitude diagram is shown in Figure \ref{pic:gaia}. The comparison color-magnitude diagram is the $<100$ parsec {\it Gaia} sample with quality cuts from Section 2.1 of \citet{Gaia2018Babusiaux}.

\subsection{Stellar parameters}

\begin{figure}
	\centering
	\includegraphics[width=1\linewidth]{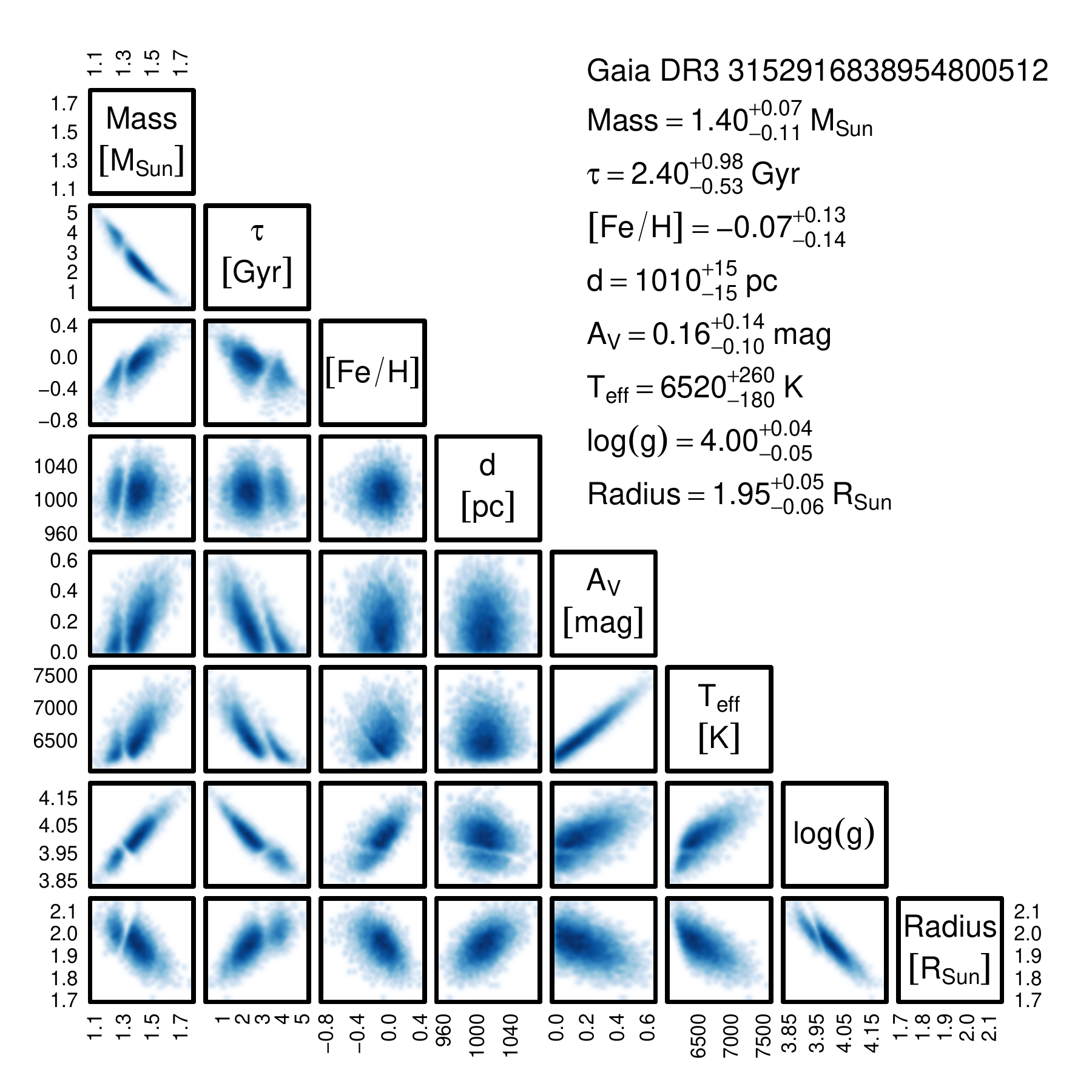}
	\caption{Best-fit stellar parameters and their posterior distributions derived from pre-occultation photometry of \jj.}
	\label{pic:mstar}
\end{figure}

We infer the fundamental and photospheric stellar parameters of Gaia
DR3 3152916838954800512 using the \texttt{isochrones} \citep{mor15}
package. Using \texttt{MultiNest} \citep{fer08,fer09,fer19}, we execute
a simultaneous Bayesian fit of the Modules for Experiments in Stellar
Evolution \citep[MESA;][]{pax11,pax13,pax18,pax19,jer23} Isochrones \&
Stellar Tracks \citep[MIST;][]{dot16,cho16} isochrone grid to a curated
collection of data for the star.  We fit the MIST grid to
\begin{enumerate}
\item
SkyMapper Southern Survey DR4 $uvgriz$ photometry including in
quadrature their zero-point uncertainties (0.03,0.02,0.01,0.01,0.01,0.02)
mag \citep{onk24};
\item
Gaia EDR3 $G$ photometry including in quadrature its zero-point uncertainty
\citep{gai16,gai21,fab21,rie21,row21,tor21};
\item
a zero point-corrected Gaia EDR3 parallax
\citep{gai21,fab21,lin21a,lin21b,row21,tor21}; and
\item
an estimated extinction value based on a three-dimensional extinction
map \citep{gree19} and the \texttt{dustmaps} Python module \citep{gre18}.
\end{enumerate}
As priors we use
\begin{enumerate}
\item
a \citet{cha03} log-normal mass prior for $M_* < 1~M_{\odot}$
joined to a \citet{sal55} power-law prior for $M_* \geq 1~M_{\odot}$;
\item
a metallicity prior based on the Geneva-Copenhagen Survey
\citep[GCS;][]{cas11};
\item
a log-uniform age prior between 1 Myr and 10 Gyr;
\item
a uniform extinction prior in the interval 0 mag $< A_{V} < 1$ mag; and
\item
a distance prior proportional to volume between the \citet{bai21}
geometric distance minus/plus five times its uncertainty.
\end{enumerate}
We plot the results of this analysis in Figure \ref{pic:mstar}.

In all following calculations, we use the following best-fit parameters from this procedure: $T_{\rm eff}=6520^{+260}_{-180}$ K, $\log g=4.00^{+0.04}_{-0.05}$, $M_*=1.40^{+0.07}_{-0.11}M_{\odot}$, $R_*=1.95^{+0.05}_{-0.06}R_{\odot}$, [Fe/H]$=-0.07^{+0.13}_{-0.14}$ dex, $A_V=0.16^{+0.14}_{-0.10}$ mag and bolometric luminosity $L_*=6.21_{-0.54}^{+0.80} L_{\odot}$. We compare these values to those from {\it Gaia} pipelines and independent reanalyses of {\it Gaia} data. Specifically, GSP-Phot pipeline\footnote{\url{https://gea.esac.esa.int/archive/documentation/GDR3/Gaia_archive/chap_datamodel/sec_dm_astrophysical_parameter_tables/ssec_dm_astrophysical_parameters.html}} derives stellar parameters from the combination of {\it Gaia} photometric data and its low resolution (BP/RP) spectroscopy. These results are quite similar to ours except the GSP-Phot metallicity is appreciably lower, [Fe/H]$=-0.51^{+0.07}_{-0.08}$ dex. \citet{zhan23} conducted a re-analysis of BP/RP spectra using data-driven approaches (rather than {\it ab initio} stellar photospheric models) and derive a near-zero extinction for \jj. There is a significant degeneracy between metallicity, extinction and temperature when the fits are made with photometry and/or low-resolution spectroscopy. Our fits presented above minimize these degeneracies by using all available photometry and the priors on extinction from the 3D maps (Ding et al. in prep.). A key conclusion for \jj\ from comparing various fitting techniques is that all inferred intervening Galactic extinction values range between $A_V=0-0.19$ mag and are well within the two-dimensional (maximum) value for \jj's direction on the sky of $E(B-V)=0.17\pm 0.02$ mag \citep{schl14}, which roughly corresponds to $A_V=0.53$ mag. For the distance of \jj, the three-dimensional map by \citet{gree19}\footnote{http://argonaut.skymaps.info/} suggests a lower value than our best-fit analysis, $E(g-r)=0.03$ mag. 

The best-fit FLAME evolutionary parameters\footnote{\url{https://gea.esac.esa.int/archive/documentation/GDR3/Data_analysis/chap_cu8par/sec_cu8par_apsis/ssec_cu8par_apsis_flame.html}} are derived from GSP-Phot measurements and the MARCS stellar photospheric models \citep{cree23}. The reported FLAME values are in reasonable agreement with our derived values. In particular, FLAME reports the evolutionary age of the star to be $\tau_*=2.0\pm 0.3$ Gyr, in agreement with our value of $\tau_*=2.40^{+0.98}_{-0.53}$ Gyr, i.e., stellar photospheric fits do not suggest that the star is young. A search for co-moving companions using methods analogous to \citet{elba21} within $10^5$ AU did not reveal any candidate co-moving stars, so there is no evidence that the star is in an open cluster.

\subsection{Galactic kinematics}

The parallax and the proper motion of the star are detected with high statistical significance in {\it Gaia}. We adopt the distance of $1.01\pm0.02$ kpc (which is a result of our full stellar photosphere fit based on the \citealt{bai21} prior), the proper motion of $\mu_{\rm RA}=-3.75\pm 0.02$ and $\mu_{\rm dec}=-7.61\pm 0.01$ mas yr$^{-1}$, and the radial velocity (RV) of $36.9\pm 3.6$ km s$^{-1}$ relative to the Solar System barycenter. A star can have a motion relative to the Sun even if it is in a purely rotational motion in the Milky Way due to projection effects and due to the peculiar motion of the Sun itself relative to the Local Standard of Rest -- these effects are reflected in the large-scale pattern in the radial velocities and proper motions measured by {\it Gaia} \citep{katz23}. We use the Galactic rotation curve from \citet{ou24}, the peculiar motion of the Sun from \citet{scho10} and the ecliptic-to-Galactic transformation equations from \citet{liu11} and \citet{reid09} to predict the motion of \jj\ relative to the Sun if it were on a circular orbit in the Galaxy, and then we calculate its peculiar velocity relative to that frame. We find that \jj\ is moving with $v_{\rm random}=37$ km s$^{-1}$, with equal contributions from the in-plane and out-of-plane motions. Such spatial velocity is not characteristic of young stars -- in fact, such velocities are more characteristic of stars with ages significantly greater than the maximal main-sequence age of a star with the mass of \jj\ \citep{Holmberg2009}. 

\subsection{Pre-occultation multi-wavelength spectral energy distribution}

\begin{figure}
	\centering
	\includegraphics[width=1\linewidth]{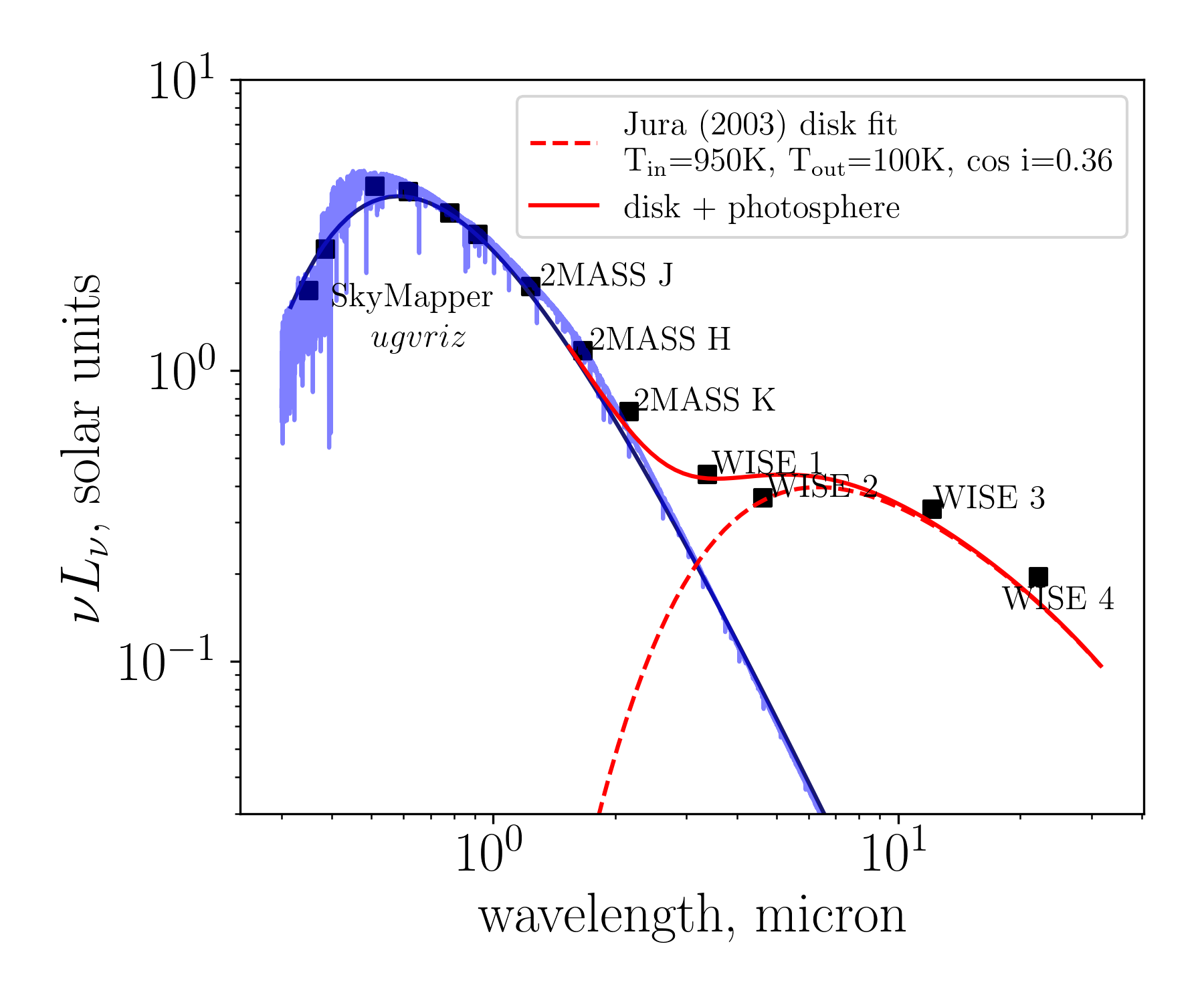}
	\caption{Photometry from SkyMapper, 2MASS and WISE. Blue: Planck spectrum (dark blue) and the BOSZ template (lighter blue) corresponding to our measured stellar parameters for \jj. Red: the best-fit disk model from \citet{jura03} with parameters as labeled (dashed line), with the full photosphere+disk fit shown in the solid line. Inclination angle $i$ is the angle between the disk axis and the observer's line of sight, so $\cos i=1$ would correspond to a face-on orientation and maximal brightness.}
	\label{pic:sed}
\end{figure}

The optical-to-infrared spectral energy distribution (SED) based on the pre-occultation photometry is shown in Figure \ref{pic:sed}. Pan-STARRS photometry is saturated at just around the brightness of our target \citep{magn13}, so we use SkyMapper Southern Survey photometry \citep{onk24}. The source is detected in the  Two Micron All Sky Survey (2MASS; \citealt{skru06}) and in the Wide-field Infrared Survey Explorer (WISE; \citealt{wrig10}) data. The source is not in the All-Sky Akari Point Source Catalog \citep{mura07} and is not covered by {\it Herschel} data. There is clearly an infrared excess well above the stellar photosphere, likely due to a presence of a circumstellar dust disk. We subtract the best-fit photospheric contribution, power-law interpolate between the observed fluxes between $2-24$\micron\ and integrate the total flux of the disk between these wavelengths. Assuming that the emission from the disk is isotropic (which is not necessarily a good assumption) and multiplying the observed flux by $4\pi d_*^2$ (where $d_*$ is the distance to the star) we obtain a disk luminosity of $L_{\rm disk}=0.63 L_{\odot}$.

A comparison of the infrared data with the best-fit optically thick, geometrically thin dust disk model from \citet{jura03} is also shown in Figure \ref{pic:sed}. In this model, the dust disk absorbs radiation from the star and re-radiates it thermally in the infrared, maintaining radiative equilibrium which sets the temperature distribution as a function of distance given by \citet{chia97}. No correction is applied for the size distribution of particles, which is a key factor for predicting or modeling longer wavelength emission \citep{woit19}. The model parameters are $\cos i$, the fraction of the area of the disk seen in projection on the sky; the inner disk temperature; the outer disk temperature; and the stellar parameters (temperature and radius) that we fix to the best-fit photospheric values for \jj. The data being fitted are the five photometric points at $\lambda\ge2$\micron, with stellar photosphere pre-subtracted so as to represent just the thermal dust emission. Using the best-fitting disk parameters, extrapolating to longer wavelengths and integrating the anisotropic emission over all viewing angles, we obtain the disk luminosity of $L_{\rm disk}=1.05 L_{\odot}$. The typical photometric uncertainty is 10\%, the level of WISE variability. Therefore, the main uncertainties in the derived disk luminosity is the uncertain contribution of the coldest gas (not well known due to lack of observations at longer wavelengths) and the uncertain degree of anisotropy. 

Our best fit slightly overpredicts the shortest wavelength infrared emission and underpredicts the longest wavelength emission, although this could be due to a known differences between the approximations by \citet{jura03} and \citet{chia97} and the full radiative transfer solution \citep{rafi06}. The outer temperature is poorly constrained by the existing data, with values $T_{\rm out}\la 150$ K yielding nearly identical fits. We have explored multi-component disks, but with only five datapoints these fits have a lot of degeneracies. The main conclusions we draw from the fits to the infrared part of the SED is that there is some evidence for both hot dust near the sublimation temperature ($900-1200$ K) and significantly colder dust, $\la 150$ K. The lack of longer wavelength data precludes us from drawing any conclusions about colder dust or about the particle size distribution which can be derived from the FIR / submm slope of the spectral energy distribution and which can inform us on whether the particles in this disk have a size distribution which is different from that of the interstellar dust \citep{banz11}. 

We construct WISE W1 and W2 lightcurves using tools described in \citet{hwan20b} and find that these fluxes vary at 0.05$-$0.1 mag level with a broad power spectrum of variability on timescales $\gtrsim 1$ day. This variability appears to be statistically significant, in that the observed power of variability is 1-2 orders of magnitude above that constructed from mock fluxes consistent with photometric uncertainties and above that constructed from reshuffled data. Phase-folding the WISE lightcurves at the low-significance peaks in the power spectrum does not reveal any periodicities. The WISE lightcurve does not cover the occultation because WISE was decommissioned in August 2024, before the beginning of the event. Pre-occultation transiting Exoplanet Survey Satellite (TESS; \citealt{rick15}) and the Zwicky Transient Facility (ZTF; \citealt{bell19}) lightcurves do not display any unusual or periodic variability. The standard deviation of the pre-occultation flux seen in TESS (where the source ID is TIC 262032775) is 0.3\%. TESS also caught the source in the middle of the occultation in late 2024, but the data are affected by scattered light, and between that and the faintness of the target during the occultation the lightcurve extraction is unreliable.  

\section{Observations during the occultation}
\label{sec:event}

\subsection{Historic and modern lightcurve}

\begin{figure*}
	\centering
	\includegraphics[width=1\linewidth]{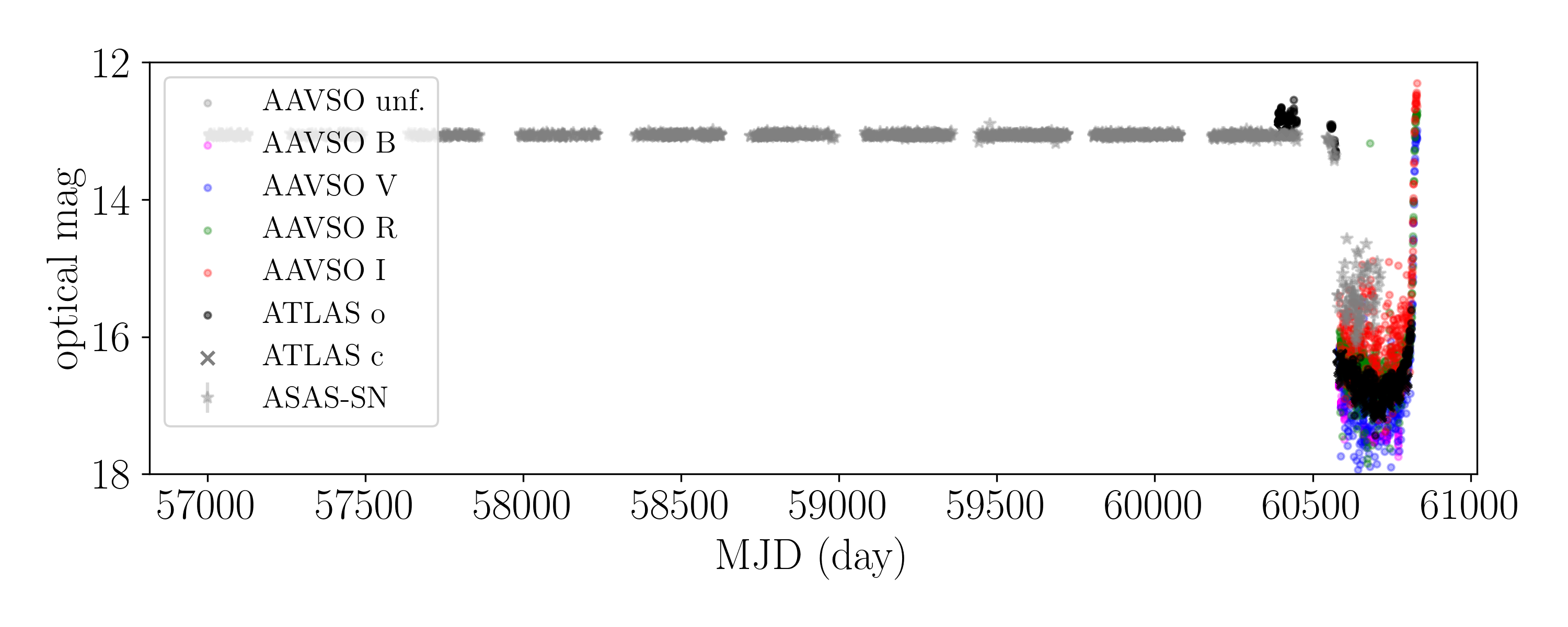}
	\includegraphics[width=1\linewidth]{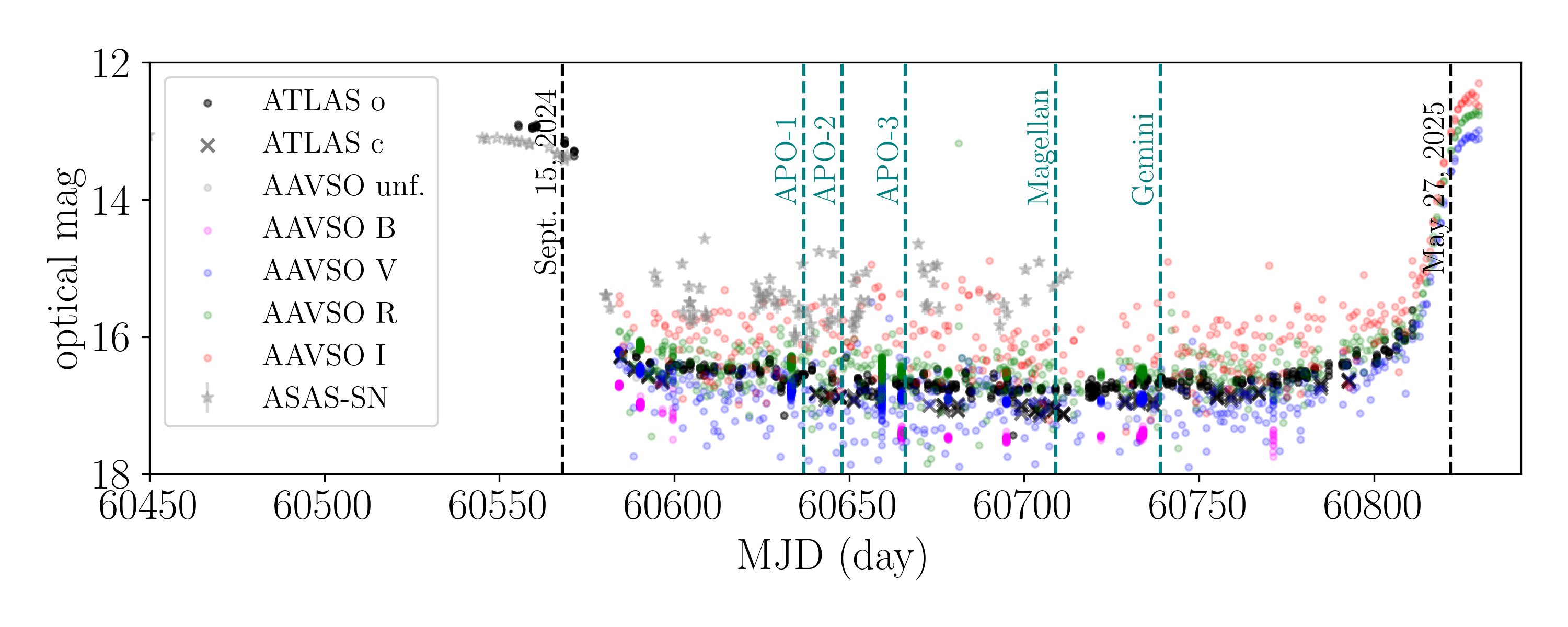}
        \includegraphics[width=1\linewidth]{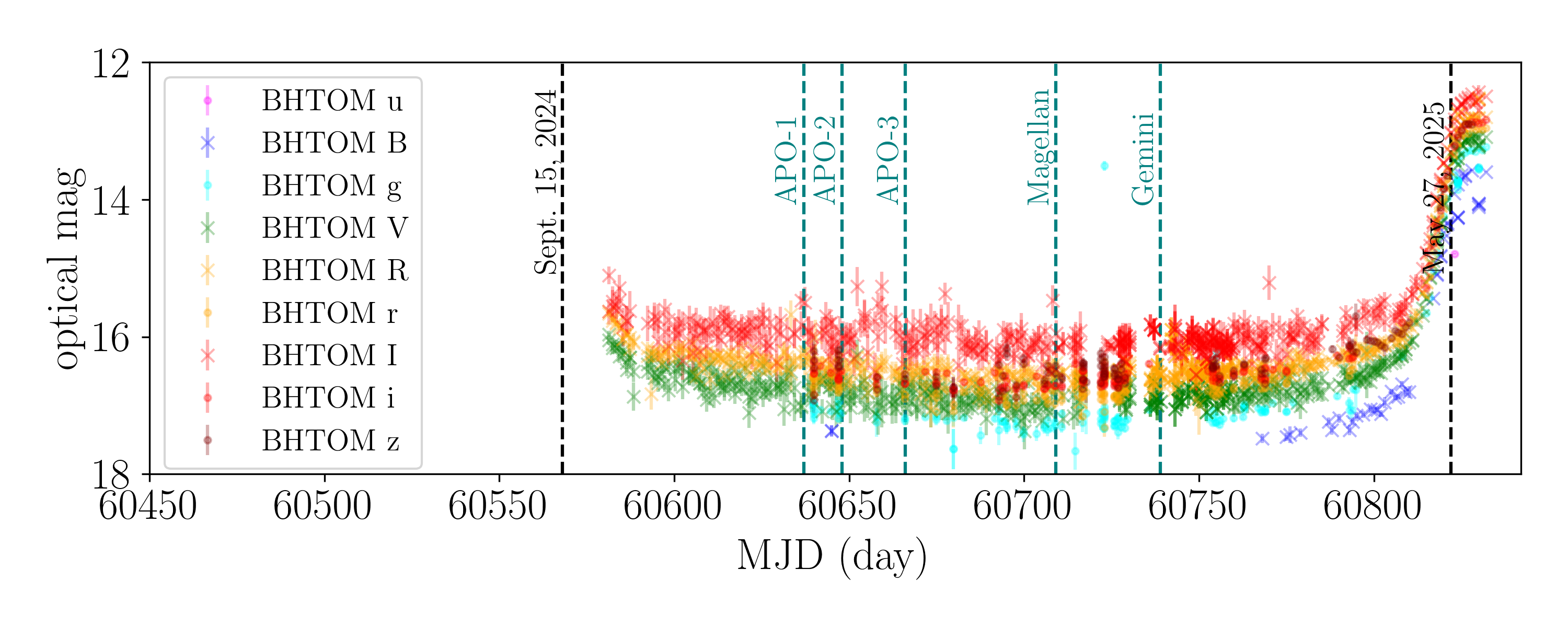}
	\caption{Top: the ASAS-SN lightcurve of \jj\ in the last 10 years. Middle and bottom: zoom in on the last few months with collection of follow-up photometry. The black dashed line at MJD=60568 (Sept 15, 2024) marks when the dimming becomes apparent in the lightcurve and the one at MJD=60822 (May 27, 2025) marks the corresponding point of the egress. Teal lines mark our follow-up spectroscopy epochs. }
	\label{pic:LC}
\end{figure*}

In Figure \ref{pic:LC} we show the lightcurve of the source over the last 10 years and over the last several months. ASAS-SN\footnote{accessed using \url{https://github.com/asas-sn/skypatrol}} \citep{koch17} conducts forced photometry at the positions of targets brighter than a certain magnitude; their pixels are 8\arcsec\ and their photometric extraction aperture radius is 2 pixels. Therefore, their photometry of dim sources can be contaminated. Indeed, searching {\it Gaia} archive within 16\arcsec\ of our target, we find 6 sources other than our target, and summing up their {\it Gaia} fluxes we find that they can accommodate 16.5 mag worth of flux within the extraction aperture. Therefore, the photometry from ASAS-SN in the dim state is likely contaminated by nearby sources. ATLAS\footnote{accessed using \url{https://fallingstar-data.com/forcedphot/}} \citep{tonr18, smit20} detects the object in both orange (o) and cyan (c) bands at 16.5$-$17th mag throughout the event. ATLAS pixels are 1.86\arcsec, so the source confusion for ATLAS is significantly less of a concern than for ASAS-SN. Nonetheless, ATLAS fluxes can be contaminated by a source 3\arcsec\ away from our target which has a nominal magnitude in {\it Gaia} of $G_{RP}=17.2$ mag, but this source looks much fainter than \jj\ in our follow-up acquisition images and therefore it is unlikely to be a concern (perhaps its {\it Gaia} photometry which was obtained during the bright state of \jj\ was unreliable). We also show the American Association of Variable Star Observers (AAVSO\footnote{accessed using \url{https://apps.aavso.org/v2/data/search/photometry/}}; \citealt{hend16, klop25}) multi-band photometry, although the uncertainty is higher than that of the ATLAS data. 

We carried out extensive multi-color photometric follow-up of the eclipse through the the Black Hole Target Observation Manager\footnote{\url{https://bhtom.space}} (BHTOM; \citealt{ziel19, wyrz20, merc25} and references therein) network of telescopes. Observations presented here were obtained on the subset of 12 telescopes involved in the network, with diameter of 0.3$-$0.9 m\footnote{Details available at \url{https://bhtom.space/public/targets/ASASSN-24fw}}. Compared to AAVSO photometry, the scatter of the data points is significantly smaller, and near-simultaneous multi-band photometry is available, enabling analysis of color evolution. Individual observers upload reduced frames (with dark, flat, and bias corrections applied), and photometric measurements are performed server-side. These are standardized using Gaia synthetic photometry \citep{mont23, merc25} on either the Johnson-Kron-Cousins system $U$, $B$, $V$, $R$, $I$, or the SDSS system $u$, $g$, $r$, $i$, $z$. 

As the source set earlier and earlier during the night, it became observable only for short periods in the evening hours, and the last ATLAS photometry available is from May 16, 2025. AAVSO and BHTOM have been able to continue observing the object in the second half of May and demonstrated a clear egress toward the bright state, with the last available observation from June 4, 2025 (AAVSO) and June 6, 2025 (BHTOM). The source will become increasingly observable again in September 2025. We take the full event to have lasted 254 days between Sept 15, 2024 and May 27, 2024, marked by vertical dashed lines in Figure \ref{pic:LC}. 

The best-sampled band is BHTOM $R$, with 609 datapoints. Even an axisymmetric structure can produce asymmetric occultations depending on its impact parameter \citep{scot14, rapp19, pram24}, but the lightcurve of \jj\ looks quite symmetric, with a formal parabolic fit to BHTOM $R$ photometry yielding a minimum at MJD=60700, $<$10 days after the mid-point of the eclipse estimated from the September 15, 2024 and May 27, 2025 start and end dates. If we force the parabolic fit to be symmetric, the residuals are at 0.1 mag level, with the first half of the eclipse slightly brighter and the second half of the eclipse slightly fainter. This slight asymmetry can be seen from the dark points in Figure \ref{pic:LC}, top. We further examine residuals from the best-fit parabolic fit to the eclipse, measure their power spectrum of variability, and compare to the power spectrum of reshuffled residuals and mock data drawn from the nominal reported photometric error. We do not find statistically significant variability relative to the best-fit smooth parabolic fit to the eclipse. Finally we split the lightcurve into two equal parts, time-reverse the second part and analyze the cross-correlation of the two chunks in search for any features like rings of gaps that would repeat after the mid-point. We find no statistically significant cross-correlation signals. 

\begin{figure}
	\centering
	\includegraphics[width=0.99\linewidth]{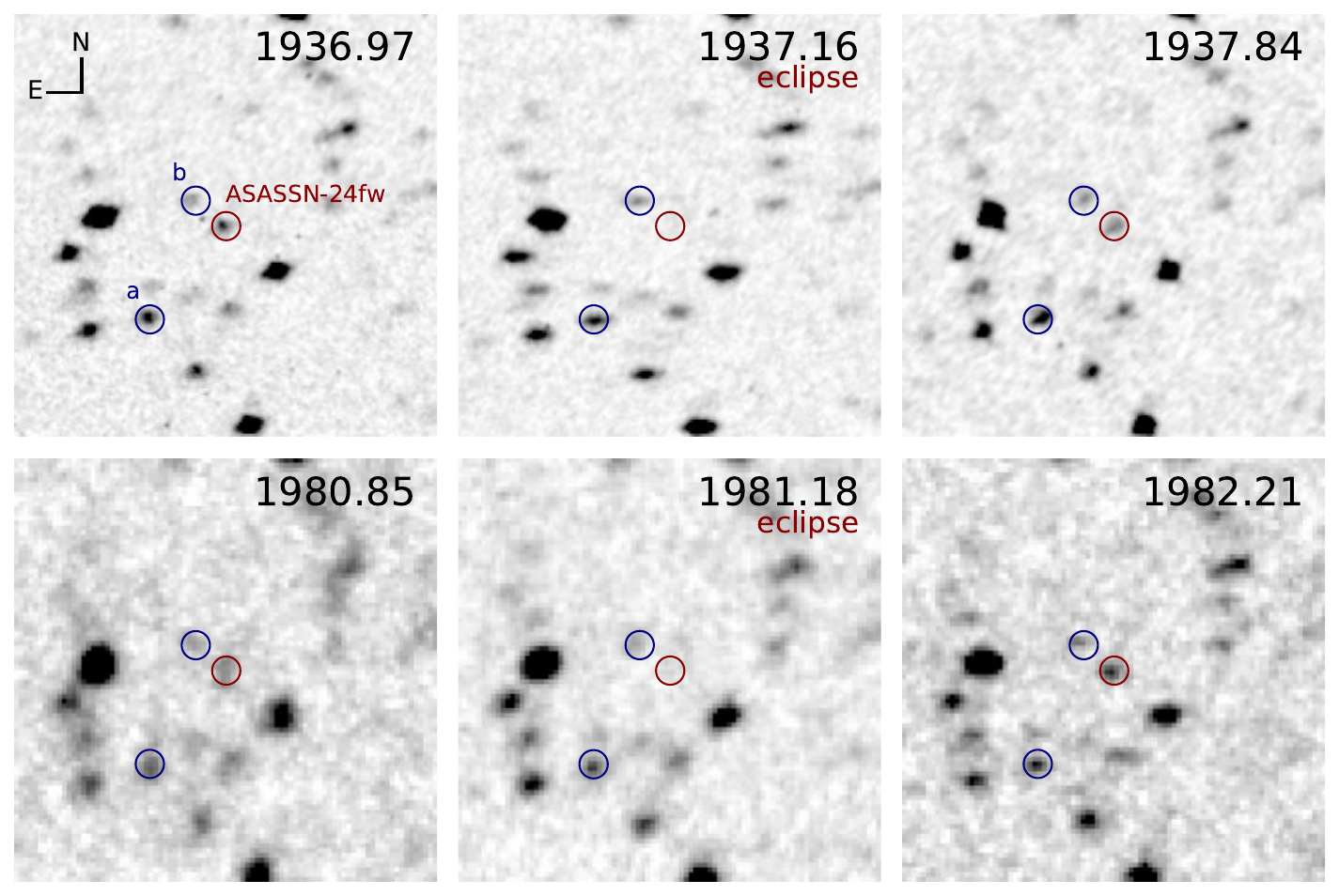}
	\caption{Before, during and after eclipse 280\arcsec$\times$280\arcsec\ snapshots from DASCH photographic plates. Top row: 1937 eclipse, bottom row: 1981 eclipse. For visual comparison, in addition to the target (\jj=ASASSN-24fw) we mark two stars of brightness similar to \jj, source a (Gaia DR3 3152915773802913792,  $G=13.01$ mag, $BP-RP$=0.70 mag) and source b (Gaia DR3 3152917212612801536, $G = 12.43$ mag; $BP-RP$=1.32 mag). Bluer source a appears brighter on photographic plates.}
	\label{pic:dasch}
\end{figure}

\begin{figure}
	\centering
	\includegraphics[width=0.99\linewidth, trim={0.7cm 0.7cm 0.5cm 0.5cm}, clip]{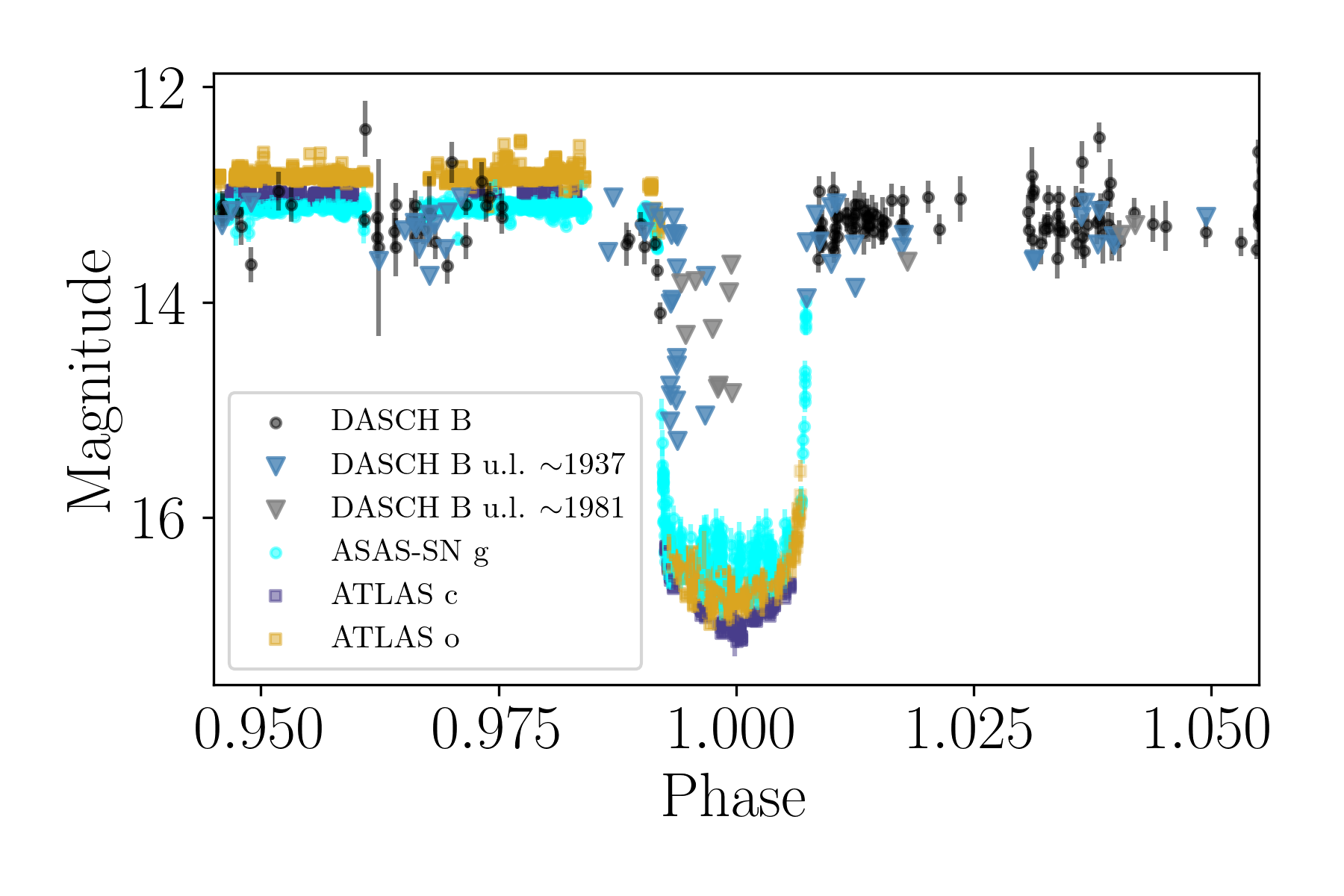}
	\caption{Historic (DASCH, including upper limits) and modern lightcurves phase-folded at $P=15,999$ days. The 1937 ingress was captured by one data point, enabling a high-precision period determination. Triangles mark DASCH upper limits.}
	\label{pic:phase}
\end{figure}

A couple of months after the start of the dimming event in \jj, \citet{nair24} conducted analysis of lightcurves from the DASCH (Digital Access to a Sky Century at Harvard) archive \citep{layc10}. They reported two earlier dimming events, one in 1937 and one in 1981, calculated the period to be 16,000 days, and predicted the end of eclipse in end of May 2025 (which turned out to be in excellent agreement with what was later observed by modern telescopes). We re-analyze the DASCH archive and confirm the findings of \citet{nair24} that \jj\ was much fainter during two previous epochs in 1937 and 1981. The 1937 ingress was caught by two plate observations, the last one at roughly 1.2 mag below the source's normal flux level. We show the in-eclipse and out-of-eclipse images from DASCH in Figure \ref{pic:dasch} to illustrate the data used to derive DASCH upper limits. With these upper limits, our best period determination is $P=15,999\pm 2$ days. The high precision is largely thanks to the recorded ingress in 1937. The phase-folded lightcurve is shown in Figure \ref{pic:phase}. The DASCH data are not sensitive enough to determine whether the shapes of the historic eclipses are the same as the one in 2024$-$25. There is no evidence that the eclipse has varied in duration. Specifically, all of the DASCH points at the phases where the object is expected to be in the occultation are upper limits, there are no detections. Outside of the occultation and within the phase range shown in Figure \ref{pic:phase}, there is only one upper limit (from 1937) that has a stated value of $>13.7$ mag, but the object is detected at 13.2 mag in five exposures within $\pm 10$ days of this upper limit. It is thus more likely that this upper limit in the un-occulted phase is due to the over-estimated sensitivity of the images rather than a true disappearance and reappearance of the source. 

We further investigate whether shorter periods ($P/2$, $P/3$ or $P/5$) could be viable alternatives. $P/2$ cannot be ruled out solely by DASCH, as there are no measurements at the expected eclipse phase, but fortunately the All-Sky Automated Survey (ASAS; \citealt{pojm97}) data cover exactly the needed time period and show no eclipse 8000 days prior to the current one. Likewise, no eclipse is seen 5333 days before the present -- DASCH and ASAS data show the star at normal brightness during those times, ruling out $P/3$. $P/4$ (or any other even integer) is ruled out since $P/2$ is excluded, and $P/5$ (or any shorter period) is inconsistent with modern survey data, e.g., ATLAS. 

\subsection{Follow-up spectroscopy}
\label{sec:followup}

We obtained five epochs of follow-up spectroscopy during the dimming event. The first three (2024-11-23, 2024-12-04, 2024-12-22) at spectral resolution $R\sim 2000$ were obtained using the 3.5m telescope at the Apache Point Observatory (APO). On November 23$-$24, 2024 (2nd half night) we obtained 75 min (on-source time) of optical spectroscopy using KOSMOS blue (3780$-$6600\AA) and 60 min red (5500$-$9630\AA) observations with slit width 1.18\arcsec\ with the position angle of the slit set along the parallactic direction. The acquisition images were manually checked against the positions of all {\it Gaia} sources in the field to ensure that the correct target was acquired. As it was cloudy and the airmass of the standard star (G 191-B2A) was not well matched to the airmass of the target, we do not attempt to calibrate these data spectrophotometrically. Arcs were obtained regularly throughout the night, so the data are as well wavelength-calibrated as is possible. The seeing was 1.0$-$1.3\arcsec. 

On December 04-05, 2024 (2nd half night) we obtained 120 min of KOSMOS blue observations with slit width 0.73\arcsec\ and the slit set to the parallactic angle, but the seeing worsened appreciably during the night from 1.4 to 2.3\arcsec, and the highest signal-to-noise combined spectrum results from the first 45 min of observations. A nearby A0V star HD52377 was observed immediately after the scientifically useful 45-min exposure, and G 191-B2A (significantly more distant on the sky from the target) was observed toward the end of the run. 

On December 22-23, 2024 (2nd half night) we obtained 112 min (on-source time) of near-infrared spectroscopy with TripleSpec (slit width 1.1\arcsec) and 45 min of KOSMOS red (slit width 0.73\arcsec) observations with slits at the parallactic angle. The seeing was 0.8\arcsec, worsening somewhat toward the end of the KOSMOS red observations. A nearby A0V star HD54102 was observed in both setups close in time to the science observations, and a more distant Feige 34 was observed at the end of KOSMOS red observations. All data reductions are performed using standard IRAF routines. All wavelength calibrations are corrected for the Earth's motion to the Solar System barycenter, and all fits are to models in air wavelengths. 

The most obvious finding from the KOSMOS spectroscopy is that of a deep Na D absorption, with equivalent widths of D$_2$ 5889.950\AA\ and D$_1$ 5895.924\AA\ lines of EW$_2=1.0\pm 0.2$\AA\ and EW$_1=0.8\pm 0.2$\AA, significantly above that expected from the stellar photosphere (EW$_2=0.36$\AA\ and EW$_1=0.31$\AA) and from interstellar contribution, given the relatively low extinction values inferred from stellar model fits to the pre-occultation photometry and {\it Gaia} low-resolution spectroscopy. We also see tentative evidence for Ca H and K (Ca II 3934\AA, 3968\AA\ doublet) emission as excess over the photospheric fits, and the equivalent width of the H$\alpha$ absorption line is significantly smaller than that of the photospheric models, suggesting that it is filled with emission. 

We therefore obtained a higher resolution $R\sim 4000$, higher signal-to-noise Magellan spectrum with MagE on 2025-02-03. We used an 0.85\arcsec\ slit and spent 1 hour on-source in four 15-minute exposures during the evening hours of the Feb 03-04 night. After the first two exposures we nodded to a nearby A0V star HD 54102 observed as a flux and additional radial velocity calibrator. All observations were made at the parallactic angle. Arcs were observed multiple times throughout the sequence. Data reductions were conducted using PypeIt \citep{proc20,pypeit} v.16.1 and moved onto the heliocentric air wavelength grid using \citet{mort91}. The most important findings from this observation include a clear detection of H$\alpha$ emission, blueshifted relative to the stellar photospheric absorption lines, complex kinematic structure within Na D absorption with multiple kinematic components apparent, and tentative evidence for Ca H and K emission. 

To probe the complex kinematic structure within the Na D absorption we requested Director's Discretionary Time on Gemini-South (GS-2025A-DD-104, PI Zakamska) with the new Gemini High-resolution Optical SpecTrograph (GHOST, \citealt{mcco24}, \citealt{kala24}). The data were obtained on 2025-03-05 and reduced using the GHOST pipeline \citep{plac24}. We used $2\times 2$ spectral and spatial binning resulting in the resolution of $R\sim 28,000$. The GHOST integral-field unit aperture is 1.2\arcsec\ on the side, whereas the nearest source to our target seen in acquisition images and in {\it Gaia} is 3\arcsec\ away, so there are no concerns of contamination by unrelated sources. The source was observed over 9 exposures, 900 sec each, for a total on-source time of 2.25 hours. The seeing varied between 0.5 and 0.6\arcsec, and the peak signal-to-noise ratio of the data was 60 (measured in the line-free and telluric-free region around 6700\AA). 

We use two BOSZ \citep{mesz20} templates\footnote{\url{https://archive.stsci.edu/hlsp/bosz}}, one closest to our best-fit solution with $T_{\rm eff}=6500$ K, $\log g=4.0$ and [Fe/H]=$0.0$ dex (hereafter high-metallicity BOSZ template) and one closest to the GSP-Phot solution with $T_{\rm eff}=6750$ K, $\log g=4.0$ and [Fe/H]=$-0.5$ dex (low-metallicity BOSZ template). We use templates with $R=2000$, 5000 and 20000 for modeling APO, Magellan and Gemini spectra, respectively. We fit several prominent absorption features in the BOSZ templates (such as Na D) to verify that in the optical range they are on the air wavelength grid. We then verify that the radial velocity of the stellar photospheric features seen in the GHOST spectrum are consistent with the {\it Gaia} value, and from then on we redshift the BOSZ templates by the {\it Gaia} value of the radial velocity in all spectral plots and modeling. 

None of the optical spectra are explicitly corrected for the telluric absorption. To identify telluric features of concern for any scientific applications, we use the set of telluric opacity curves \verb|telfit_maunakea_3100_26100_r20000.fits| from the PypeIt distribution \citep{proc20} corrected to air wavelengths and to the heliocentric frame. We find that individual molecular absorption features in the red part of the spectrum line up perfectly with those seen in the GHOST data, so the velocities are well calibrated. For example, in the wavelength range $6880-6920$\AA, where there are 18 strong well-separated telluric features and relatively weak stellar photospheric features, the velocity offset between the data and the telluric template is $0.05\pm 0.11$ km s$^{-1}$. From these fits we obtain the spectral resolution of the GHOST data, $\sigma_v=3.8$ km s$^{-1}$, assuming that it dominates the observed widths of the telluric features. We show the telluric template with the strongest absorption on all figures to guide the eye. For example, for the wavelength range around Na D absorption, there are identifiable telluric features with peak strength of a few per cent of the continuum, comparable to the rms noise in the data. They affect the $\chi^2$ of the fits but not any scientific conclusions. In the red part of the spectrum telluric absorption reaches optically thick levels, and we avoid using any features in this part of the spectrum. 

Because of the superb quality of the GHOST data, in what follows we primarily present the results of this observation. The other spectroscopic epochs provide important information on the variability of the observed features throughout the dimming event, so we discuss the comparison between the GHOST observation and the previous observations as needed. 

GHOST spectrum, though impacted by the occulter, offers information on the stellar photosphere. Photospheric absorption lines can be contaminated by gas absorption and gas emission described below and by the faint binary companion. We select 17 relatively strong (peak amplitude relative to the continuum flux density of 0.3 or above), isolated (no other strong lines in the BOSZ template within 1\AA) absorption lines with no signs of gaseous emission and toward the blue part of the spectrum where the contamination from telluric absorption and from the putative low-mass companion is negligible. When fitted with Gaussians, their velocity dispersion is $\sigma_v=15\pm 3$ km s$^{-1}$ (mean and sample standard deviation), so they are spectrally resolved in the GHOST spectrum. We then fit them as an ensemble with the stellar rotational profile from \citet{diaz11} and obtain $v_{\rm rot}\sin i=32\pm 3$ km s$^{-1}$, consistent with the inferred mass and age of \jj\ \citep{amar19}. Given the quality of the data, there is no statistical preference between Gaussian and \citet{diaz11} rotational profiles, nor any need for more complex blended profiles \citep{hoad20}. In what follows we explicitly state when the templates have been rotationally broadened using Gaussian broadening.

Li I 6707.76\AA, 6707.91\AA\ doublet can serve as an indicator of stellar youth, as primordial lithium is easily destroyed by proton captures in the convective envelope of a newly formed star and is not replenished by stellar nuclear reactions \citep{stau98}. No Li I absorption is apparent in the GHOST spectrum of \jj. To obtain a formal upper limit, we fix the velocity centroids of the features to the value expected for the {\it Gaia} radial velocity and the velocity dispersion to $\sigma_v=15$ km s$^{-1}$ as calculated above. We obtain a nominal 3$\sigma$ upper limit on the total equivalent width of the lithium doublet of 54 m\AA. Examining spectral features with similar wavelengths demonstrates that 50 m\AA\ absorption lines are quite easily seen in the spectrum, so it is likely that the upper limit is conservative and Li I with EW$=30$ m\AA\ would have been at least tentatively seen, and it is not. An upper limit of EW$=30-50$ m\AA\ implies a lower limit of 130 Myr for a star with an effective temperature of 6500 K (\citealt{jeff23}, at the edge of their available temperature grid). \citet{guti24} demonstrate that all open cluster stars at $T_{\rm eff}=6500$ K with EW$=30$ m\AA\ and below have ages of several hundred Myr. Although for relatively hot F stars there is no one-to-one relationship between age and Li I EW \citep{sode10}, these recent models and data suggest that \jj\ is not a young star.

\subsection{Sodium absorption}
\label{sec:nad}

\begin{figure}
	\centering
	\includegraphics[width=1\linewidth]{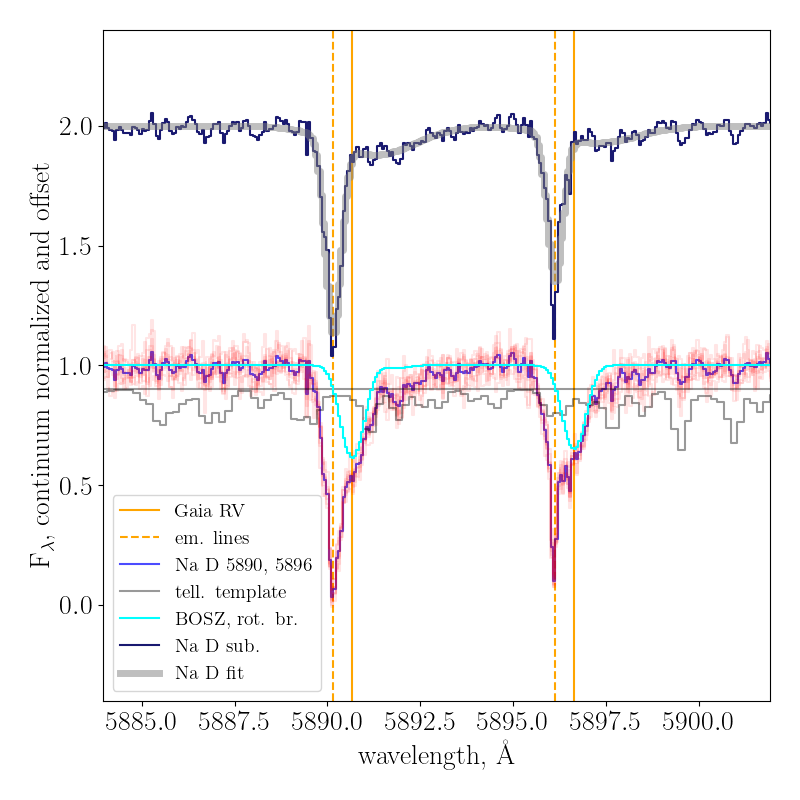}
	\caption{Top: the GHOST spectrum around Na D $\lambda\lambda$ 5889.950, 5895.924\AA\ doublet in the heliocentric frame, with the stellar photosphere divided out. The thick grey line shows a 3-Gaussian fit. Middle: 9 individual 15-minute spectra (pink) and the mean-combined spectrum (blue) normalized to be 1 in the continuum as observed, including the stellar photosphere. Solid cyan line shows the rotationally broadened BOSZ template shifted to the stellar frame based on the {\it Gaia} radial velocity, excluding the Ni I 5892.88\AA\ line not seen in our source. Bottom grey curve shows a telluric absorption template to demonstrate which of the small features in the observed spectrum could be telluric. The solid orange vertical line shows centroid of the photospheric absorption lines expected based on the {\it Gaia} radial velocity, and the orange dashed line shows the centroid of the metal emission lines.}
	\label{pic:ghost_na}
\end{figure}

\begin{figure}
	\centering
	\includegraphics[width=1\linewidth]{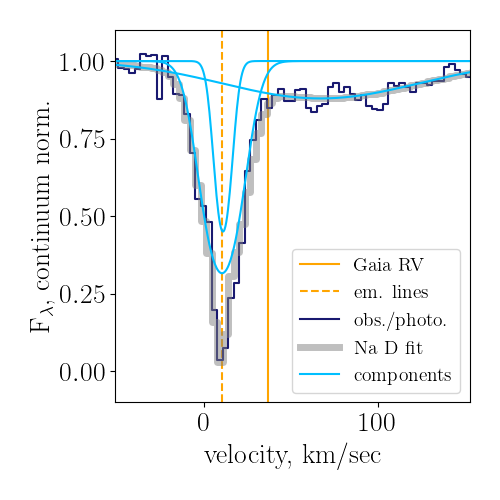}
	\caption{The individual kinematic components of $\tau_{\rm occ}(\lambda)$ (light blue) and the overall fit (thick grey) to the photosphere-corrected Na D $\lambda$ 5889.950\AA\ line in the GHOST spectrum as a function of the heliocentric velocity. As in other figures, the solid orange vertical line shows the centroid of the photospheric absorption lines expected at the {\it Gaia} radial velocity, and the orange dashed line shows the centroid of the metal emission lines.}
	\label{pic:ghost_na_small}
\end{figure}

All follow-up spectra show deep Na I D absorption, inconsistent with the stellar photosphere, indicating that there is a large column density of neutral gas in front of the star. The observed GHOST spectrum of the Na I D doublet (pink and purple), the observed spectrum corrected for the photospheric contribution (dark blue), and the fit comprised of multiple Gaussian components (thick grey line) are shown in Figures \ref{pic:ghost_na} and \ref{pic:ghost_na_small}. In this modeling, we assume that the observed spectrum, when normalized to have local continuum $=1$, is described by 
\begin{equation}
F_{{\rm obs}, \lambda}=F_{{\rm ph}, \lambda} \exp(-\tau_{\rm occ}(\lambda)), 
\label{eq:NaD}
\end{equation} 
where $F_{{\rm ph}, \lambda}$ is the photospheric spectrum before it is affected by the gas absorption and $\tau_{\rm occ}(\lambda)$ is the opacity due to the occulter. Therefore, to derive the gas opacity in the occulter we must first determine the underlying photospheric spectrum. We correct for the photospheric contribution to Na I D using a variety of methods. We extract Na I D photospheric opacities from the BOSZ templates over the wavelength of interest by fitting single Gaussian profiles to them and then rotationally broaden them either using the Gaussian profile or the \citet{diaz11} profile with the parameters derived in Sec. \ref{sec:followup}. In this rotationally-broadened model spectrum $F_{{\rm ph}, \lambda}$, we either match the equivalent widths of Na I D lines to those of BOSZ templates or the wavelength-integrated opacities. The differences in these methods do not alter the final measured equivalent widths of Na I D by more than 5\%. The top curve in Figure \ref{pic:ghost_na} shows the result of this procedure, i.e., $F_{{\rm obs}, \lambda}/F_{{\rm ph}, \lambda}$, which should be just the result of absorption by sodium in the occulter.

Having corrected for the photospheric absorption, we then fit the Na D wavelength-dependent opacity $\tau_{\rm occ}(\lambda)$ due to the occulter with two or three Gaussian components. We restrict both lines within the Na D doublet to have the same kinematic substructure -- i.e., we require that the centroid velocities and the velocity dispersions of all Gaussians are the same for both lines, but we do not require their amplitudes to be fixed at any particular ratio. The Gaussian functions do not necessarily represent the true underlying kinematics of the gas, plus the fit has some sensitivity to the initial parameters, to the assumed bounds on the parameters, to the exact procedure for photospheric correction, and to the choice of procedure for continuum normalization, but some salient features of the fit are quite robust. In particular, the fit with three components is significantly better than with two in all cases that we tried. The narrowest component ($\sigma_1=4.5$ km s$^{-1}$, $v_1=10$ km s$^{-1}$, i.e., offset by $-27$ km s$^{-1}$ from the systemic velocity of the star) is spectrally unresolved at the GHOST resolution. Its dispersion and centroid vary by only 1 km s$^{-1}$ depending on the photospheric correction and any other assumptions. The second narrowest component is also significantly blueshifted compared to the systemic velocity of the star ($\sigma_2=10-13$ km s$^{-1}$ and $v_2=10-12$ km s$^{-1}$). These two components account for 40\% each of the integrated opacity of the Na D absorption. The broadest component (the remaining 20\%) is less robust to the exact assumptions of the fitting model ($\sigma_3=40-55$ km s$^{-1}$ and $v_3=55-75$ km s$^{-1}$, i.e., redshifted by $20-40$ km s$^{-1}$ relative to the systemic velocity), but is nonetheless required in all fits (in a typical case, reduced $\chi^2$ decreases from 2.6 to 1.6). Importantly, the Magellan MagE spectrum, while it does not have a high enough resolution to separate the two narrowest components, shows the same overall kinematic structure of the Na D lines, with a clear redshifted wing accompanying a stronger blueshifted component. 

The ratio of the velocity-integrated opacity of Na I D$_2$ (5889.950\AA) to that of D$_1$ (5895.924\AA) is 1.7$-$1.9, close to the quantum-mechanical value of 2:1. The overall equivalent widths of absorption (excluding the photospheric component) for the fit shown in Figures \ref{pic:ghost_na} and \ref{pic:ghost_na_small} are EW$_2=0.74\pm 0.07$\AA\ and EW$_1=0.50\pm 0.05$\AA. The uncertainty is dominated by which wavelength ranges are chosen to normalize the overall shape of the continuum. The results shown in Figure \ref{pic:ghost_na} are from median-averaging observed flux densities at wavelengths $5878-5880$\AA\ and $5906-5908$\AA, linearly interpolating between these values, and dividing the spectrum by this interpolated continuum.

An important question is whether any of the kinematic components of Na D can be due to the intervening Galactic gas, especially the narrowest component which is spectrally unresolved in the GHOST spectrum. The equivalent widths of this component alone are 0.13-0.16\AA\ and 0.05-0.11\AA\ (depending on the fitting method and photospheric correction) for D$_2$ and D$_1$, respectively. The fractional errors on the opacities of individual components of the fit are significantly higher than those on the overall equivalent width due to degeneracies between components, but the narrow component is required in all fits we have tried, and within the uncertainties the ratio of opacities of narrow D$_2$ to narrow D$_1$ is consistent with the quantum-mechanical value 2:1. According to the relationships of \citet{pozn12} and \citet{murg15}, such equivalent widths can arise in the interstellar medium that produces extinction values $E(B-V)\simeq 0.05$ mag, or $A_V\simeq 0.15$ mag. The extinction values derived from the pre-occultation {\it Gaia} data for our target range between $A_V=0$ and $0.19$ mag depending on the exact fitting approach (our best fit is $A_V=0.16$ mag), so the narrowest component may arise in the intervening interstellar medium between us and \jj, and this would be consistent with the (uncertain) reddening measurement.

However, the kinematics of the narrowest component is somewhat inconsistent with the expected Galactic rotational motion of the interstellar gas along the line of sight to \jj. When we look in the direction of \jj\ ($l\simeq 210$ degrees) from the location of the Sun, we find that the Galactic rotation curve results in a net redshift of all objects along the line of sight as is apparent in the {\it Gaia} radial velocities map of the Galaxy \citep{katz23}. In addition to the Galactic rotation, we need to account for the peculiar motion of the Sun relative to the Local standard of rest. With the best available values of $U_{\odot}=11.1$ km s$^{-1}$ (toward the center of the Galaxy) and $V_{\odot}=12.2$ km s$^{-1}$ (faster than the local rotation curve) \citep{scho10}, the Sun's motion is essentially anti-parallel to the direction toward \jj, and therefore the minimal redshift of the interstellar gas that follows Galactic rotation is expected to be $15.6$ km s$^{-1}$, more than the observed redshift of the narrow components of $10$ km s$^{-1}$. While this difference in velocity does not completely rule out interstellar Na D absorption, in what follows we proceed with the assumption that none of the observed Na D absorption structure is due to the intervening interstellar medium. 

Debris disks with cold gas (CO) detections commonly show Ca II H K and Na I D absorption in their spectra \citep{igle18, rebo18}, likely due to the hot gas in the vicinity of the star where it may have been generated during the photoevaporation of solid byproducts of planet formation \citep{choq17}. The equivalent width of the resulting absorption is one-two orders of magnitude below that seen in \jj\ \citep{rebo18}, but the circumstellar disk in \jj\ is also significantly more luminous than theirs. In what follows we assume that Na D is associated with the occulter rather than in the circumstellar material. Both the interstellar absorption and the circumstellar absorption should still be there after the occultation has concluded. Therefore, our assumptions about the origin of the Na I D absorption exclusively in the occulter can be tested by a high-resolution post-occultation spectrum which can be obtained after the source is visible again in Fall 2025. 

\citet{fore25} analyze the spectral energy distribution of \jj\ during the occultation and hypothesize that the near-infrared excess may be due to an M-dwarf companion which contributes $\sim 20\%$ of the flux at 1\micron. Their analysis provides the current best upper limit on the luminosity of the companion if some of the near-infrared emission seen during the dimming is due to hot dust. Taking a BOSZ template with $T_{\rm eff}=3500$ K and normalizing it to the maximum value allowed by their SED analysis, we find that during the occultation the companion contributes $\sim 1\%$ of flux at the wavelength of Na D, with a similar equivalent width. Therefore if the companion is kinematically offset from the brighter star it may alter the measured kinematics of Na D at 1\% level but not qualitatively change our results.

\subsection{Hydrogen emission}

\begin{figure}
	\centering
	\includegraphics[width=1\linewidth]{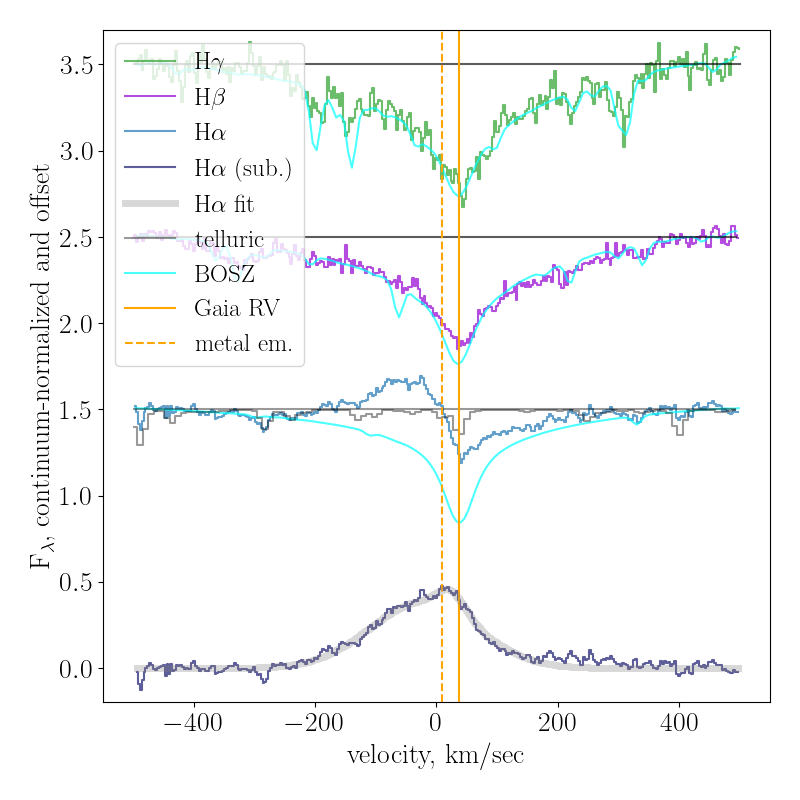}
	\caption{Balmer series lines H$\alpha$, H$\beta$, and H$\gamma$ as seen by GHOST compared with the low-metallicity BOSZ templates. The bottom curve shows H$\alpha$ after subtraction of the BOSZ template and a two-Gaussian fit to the emission profile. No rotational broadening was applied to the template as these lines are intrinsically} broad.
	\label{pic:halpha}
\end{figure}

In the GHOST spectrum, H$\alpha$ line shows broad ($\sim 200$ km s$^{-1}$) emission which has a net blueshift relative to the star of $\sim 50$ km s$^{-1}$ and an asymmetric velocity profile displaying blue excess (Fig. \ref{pic:halpha}). There is some evidence for small amounts of H$\beta$ emission only detectable in the deviations of the observed velocity profile of H$\beta$ from that in the BOSZ template (Fig. \ref{pic:halpha}). There is no evidence for emission in the higher-order Balmer lines. 

A two-Gaussian fit to H$\alpha$ yields an equivalent width measurement of EW(H$\alpha$)=1.98$\pm$0.02\AA\ (statistical fitting uncertainty only). Since the line is blended, with no obvious individual components, the individual parameters of the fit -- such as the Gaussian centroids and dispersions -- are degenerate with one another, but the overall equivalent width is quite robust, with alternative fitting models yielding the same value. The net flux-weighted velocity centroid is at $-19\pm 3$ km s$^{-1}$ from the heliocentric frame, or at $-56\pm 4$ km s$^{-1}$ with respect to the stellar frame. A single-Gaussian fit yields an effective velocity dispersion of 84 km s$^{-1}$. 

\begin{figure}
	\centering
	\includegraphics[width=1\linewidth]{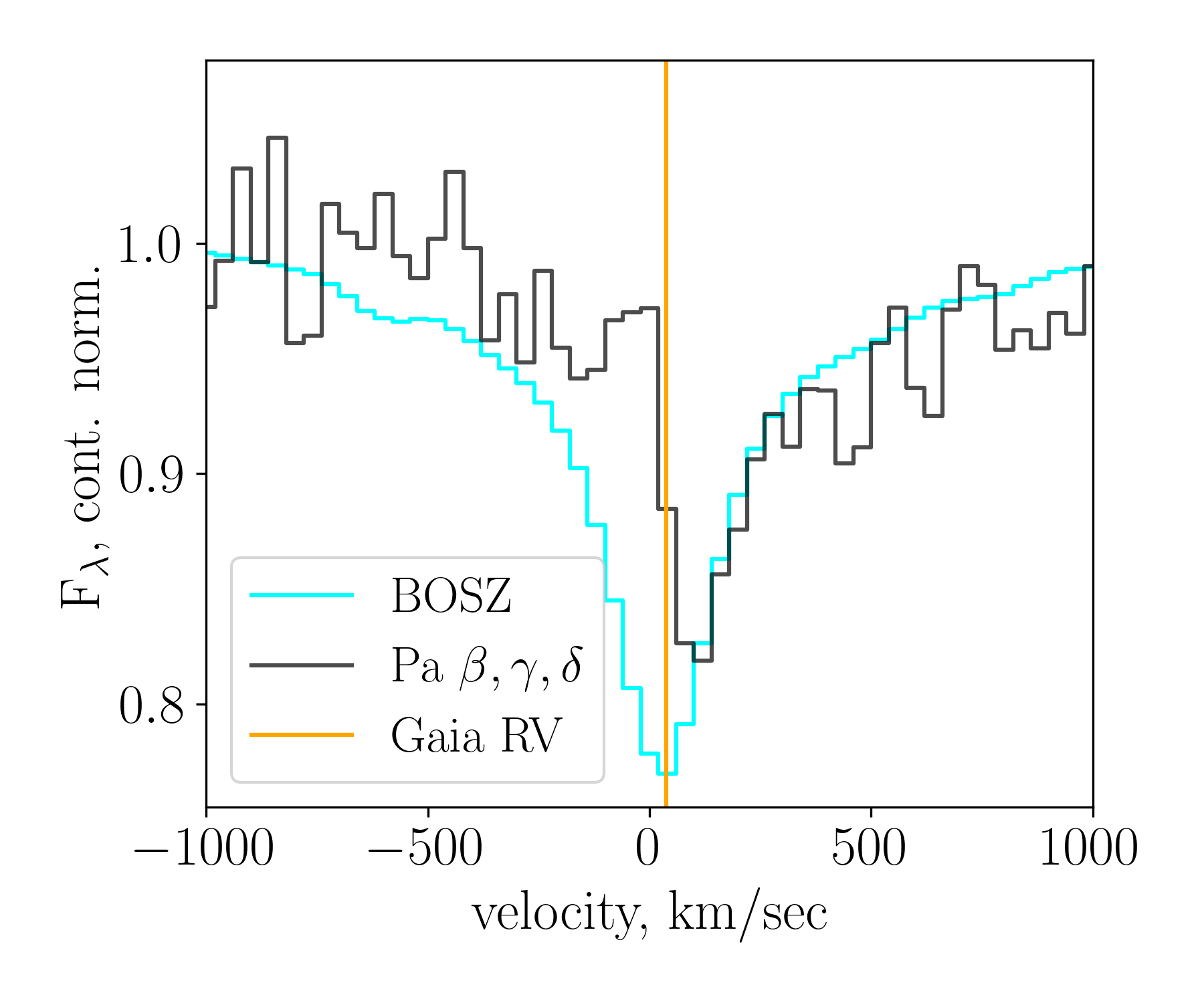}
	\caption{Stacked and continuum-normalized data and BOSZ low-metallicity template for Paschen $\beta$, $\gamma$ and $\delta$ lines in the APO TripleSpec data. All show emission-line excess individually.}
	\label{pic:paschen}
\end{figure}

The TripleSpec near-infrared observation in the 3rd spectroscopic epoch does provide additional information, which is that emission line excess is detected in Pa$\beta$, Pa$\gamma$ and Pa$\delta$, i.e., all strong hydrogen lines that fall into the $J$ band that are not affected by the telluric absorption. To slightly improve signal-to-noise ratio, in Figure \ref{pic:paschen} we show the stacked line profile for these three lines from TripleSpec data in comparison with the similarly stacked BOSZ model; the velocity scale is heliocentric as in all other figures. The kinematics of this profile is consistent with that seen in the H$\alpha$ line. 

Photospheric fitting, lithium non-detection and Galactic kinematics suggest that \jj\ is over 2 Gyr old, yet it has both an infrared excess and H$\alpha$ emission. Strong H$\alpha$ emission is commonly seen in young stellar objects with a diversity of shapes which are due to the optical depth effects and the geometry of stellar gas accretion and winds. In the atlas of high-resolution spectra of H$\alpha$ in young stellar objects by \citet{reip96}, the shape of H$\alpha$ in \jj\ would be characterized as type IV-R (one-sided blue excess with the red part of the photospheric absorption line profile visible), and a handful of objects in their catalog display such shape. The width of H$\alpha$ in \jj\ is well within the range of velocity widths seen in young stellar objects and well within the critical rotational velocity of the star $\sim$370 km s$^{-1}$. What is clearly different in \jj\ is the strength of H$\alpha$: the median equivalent width in the \citet{reip96} sample is 35\AA, and the distribution extends to hundreds of \AA. The relative paucity of H$\alpha$ emission in \jj\ is consistent with it being significantly older than the sources in \citet{reip96}. Stars which host debris disks do not always show H$\alpha$ emission \citep{curr08}. The equivalent widths of a few \AA\ -- similar to that seen in \jj\ -- are indeed typical, though unlike \jj\ the stars themselves still often show additional signs of youth, e.g., strong reddening due to circumstellar dust \citep{curr08}. 

A putative M-dwarf companion described by \citet{fore25} would contribute $\sim 3\%$ of the continuum flux around H$\alpha$ and up to $\sim 30\%$ of the continuum flux around Pa$\beta$, but would not have hydrogen absorption in its spectrum, so its effect would be to reduce the equivalent width of the stellar photospheric absorption features somewhat, and therefore our measurement of the equivalent width of emission can be under-estimated. 

A blue-shifted, broad H$\alpha$ emission is seen in all our five epochs of spectroscopy with some variability, with EW$=1.4\pm0.3$\AA\ in the APO1 data, $1.2\pm0.2$\AA\ in the APO2 data, $2.3\pm 0.4$\AA\ in the APO3 data and $1.8\pm0.2$\AA\ in the Magellan data. The velocity centroids are consistent with the GHOST value within their respective uncertainties. But a by-eye examination of the spectra presented by \citet{fore25} reveal a possibility of much more dramatic variability, with H$\alpha$ emission barely, if at all, detected in their October 2024 spectrum, and the shape of Paschen lines appearing to change between the two epochs of their observation. This possible variability will need to be taken into account in any interpretation of the post-occultation spectra.

In what follows we assume that the H$\alpha$ emission is due to circumstellar activity, e.g., magnetospheric accretion onto the F star. This assumption can be tested using post-occultation spectroscopy: if the emission originates close to the star on scales much smaller than the occulter, it should still be there with the same equivalent width. If the emission is on the same scales as the occulter, it may emerge with different kinematics \citep{buda05}, and if it originates in the occulter itself it would retain the kinematics, but appear with a much lower equivalent width. The velocity width of H$\alpha$ observed in \jj\ is about twice that of the H$\alpha$ emission seen in low-mass T Tauri stars \citep{whit03}. Furthermore, if the H$\alpha$ emission were associated with the putative M-dwarf companion, then its intrinsic equivalent width (undiluted by the light from the F star) would be 60\AA, in the upper 20\% for such stars \citep{whit03}. While these two arguments do not completely rule out the possibility that hydrogen emission is associated with the occulter, they do make it somewhat unlikely.

\subsection{Metal emission}

\begin{figure*}
	\centering
	\includegraphics[width=0.32\linewidth]{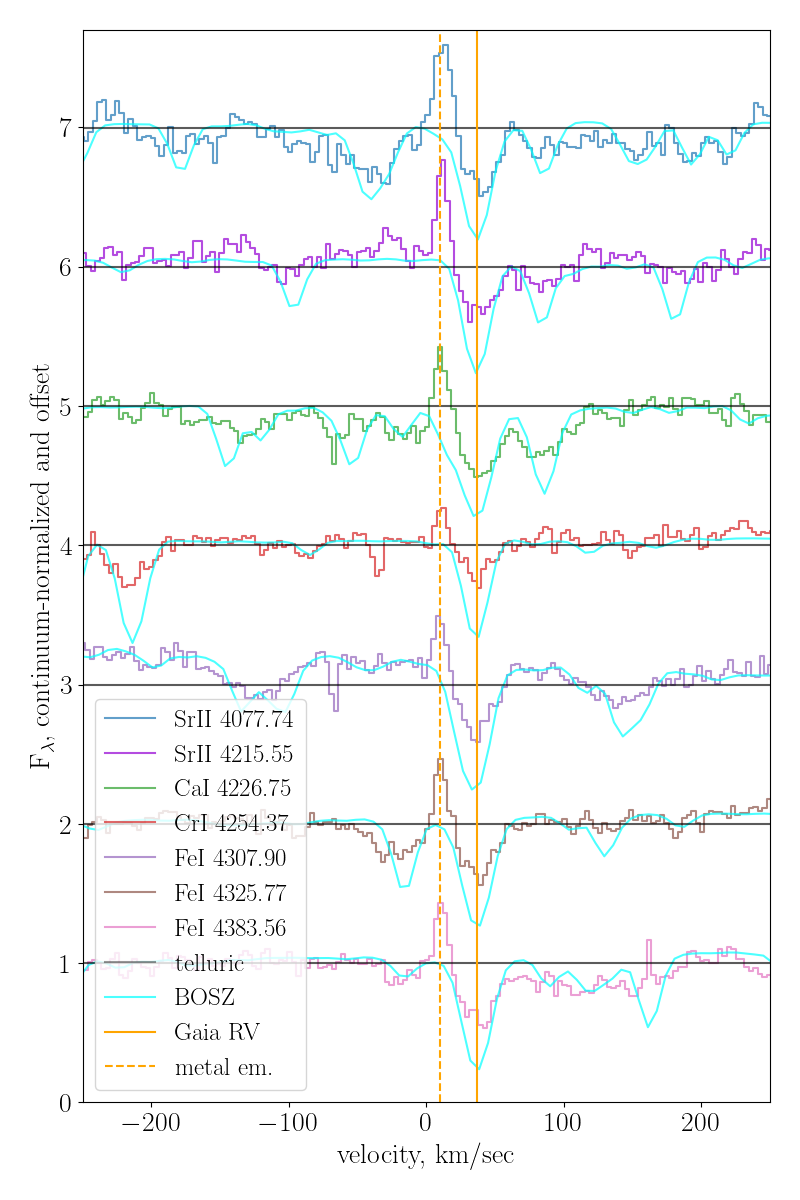}
	\includegraphics[width=0.32\linewidth]{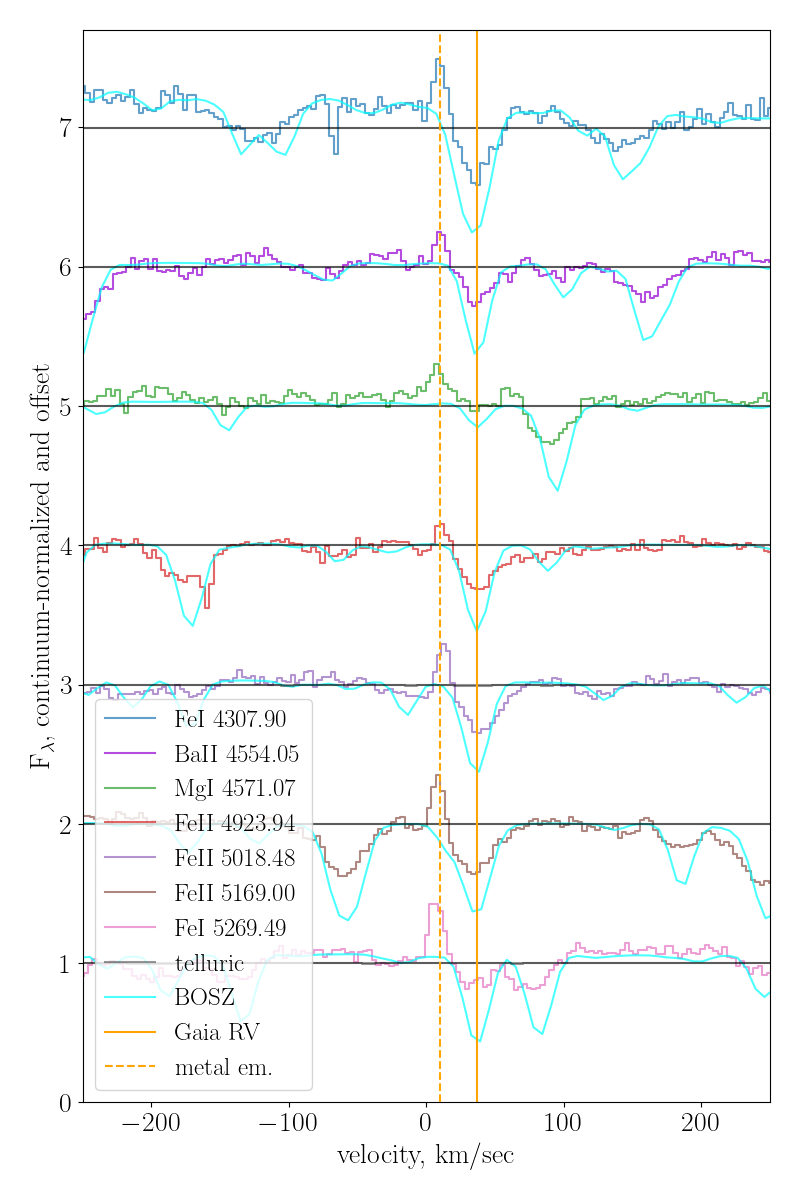}
	\includegraphics[width=0.32\linewidth]{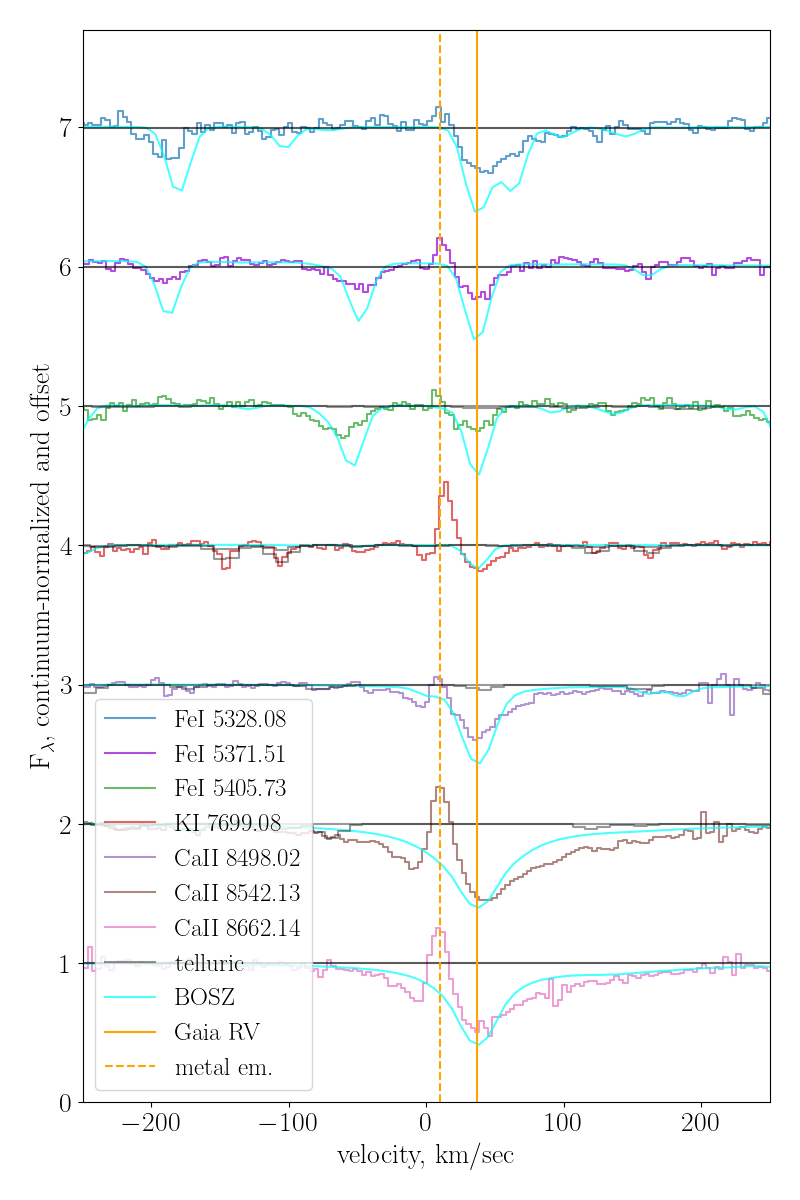}
	\caption{All metal emission lines in the GHOST spectrum with confident NIST identifications, organized from shortest to longest wavelengths. The observed spectrum is shown in the heliocentric frame. Comparison BOSZ low-metallicity templates (cyan) are shown shifted by the {\it Gaia} RV of 37 km s$^{-1}$ and with no rotational broadening applied. Telluric templates are shown in grey for comparison. Solid orange line shows the systemic velocity of the star (37 km s$^{-1}$) and dashed orange line shows the typical velocity centroid of the metal emission lines (10 km s$^{-1}$).}
	\label{pic:metals}
\end{figure*}

There are some hints of Ca H and K emission in KOSMOS and Magellan spectra, detectable only as a subtle excess over the BOSZ templates overplotted on the spectra at a matched spectral resolution. While GHOST data over Ca H and K are too low signal-to-noise for a definitive measurement, a striking new finding in the GHOST spectra is that of a couple of dozen metal emission lines (Fig. \ref{pic:metals}). We split the entire GHOST spectrum into 1000 km s$^{-1}$ chunks with comparison telluric and BOSZ templates overplotted and we systematically examine all of them by eye. Neither the low-metallicity BOSZ template nor the high-metallicity one faithfully represent the equivalent widths of many of the stellar photospheric features. In general the low-metallicity, higher temperature BOSZ template shows lower equivalent widths of absorption features which are in better agreement with the data, but for example Ca II IRT lines (the Ca II 8492\AA, 8542\AA, and 8662\AA\ triplet) are much better represented by the high-metallicity, lower temperature BOSZ template. It is possible that there are abundance patterns not captured by the [Fe/H] metallicity indicator. Abundances of individual elements, stellar rotation that broadens the features compared to the template, and the emission that fills in the stellar absorption features are challenging to disentangle. In what follows, we explicitly state which templates are used in the figures. 

We make a list of confident and tentative emission-line detections. We then fit each feature with two Gaussians, one to account for the emission and one to account for the stellar photospheric absorption, and we obtain the wavelength of each emission line. We then examine NIST within 0.3\AA\ (which is 15 km s$^{-1}$ at 6000\AA) of each measured line centroid and find the best identification, preferring close wavelength match, common elements with relatively low-energy upper states and higher transition probabilities when in doubt. 

The equivalent widths of the features shown in Fig. \ref{pic:metals} range between 0.03\AA\ (for Mg I 4571\AA) to 1.3\AA\ (for Ca II 8542\AA). All emission lines have kinematics similar to each other: the mean centroid velocity is 10.7 km s$^{-1}$ with a standard deviation within the sample of 21 lines of only 1.7 km s$^{-1}$. The fitted velocity dispersions range between 4$-$8 km s$^{-1}$, with most values around 5 km s$^{-1}$, i.e., close to the limit of the GHOST resolution. The kinematics of the metal emission lines is similar to that of the narrow components of Na D absorption, suggesting common origin.

\citet{moto13} present a high-resolution study of the Ca II triplet in young stellar objects, transitional disks, and young main-sequence stars. Their sample shows Ca II 8542\AA\ emission with equivalent width up to an order of magnitude above that seen in \jj. On average, the equivalent width of emission declines as a function of the evolutionary stage, so it is not out of question that Ca II emission of the strength seen in \jj\ could originate in the inner regions of the circumstellar disk, especially given that there are other indications of the gas presence in the inner circumstellar disk (i.e., H$\alpha$). The net velocity offset between Ca II emission and the stellar photosphere seen in our source -- blueshifted by $27$ km s$^{-1}$ -- is on the high end of the distribution of radial velocities seen by \citet{moto13}, but not unprecendented. What is strikingly different about \jj\ compared to the young and transitional stellar objects in their sample is the velocity dispersion: their measured median velocity dispersion is $\sim 45$ km s$^{-1}$, with the lowest value of 23 km s$^{-1}$ (their spectral resolution is higher than ours). These values make sense for gas that probes circumstellar dynamics. In contrast, the emission peaks in \jj\ seen in Figure \ref{pic:metals} have a velocity dispersion of $<10$ km s$^{-1}$ and some are spectrally unresolved even at the spectral resolution of GHOST ($\sim 4$ km s$^{-1}$). 

\begin{figure}
	\centering
	\includegraphics[width=1\linewidth]{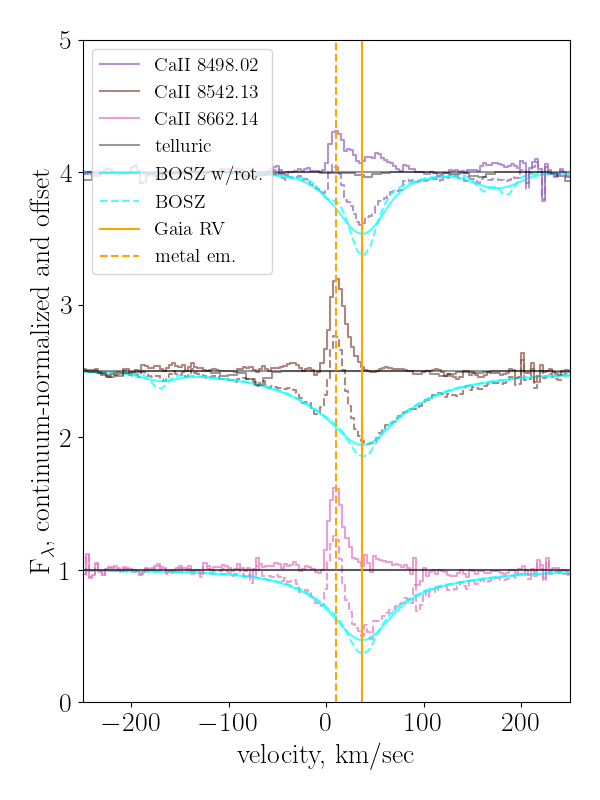}
	\caption{Calcium triplet in the GHOST spectrum shown with the high-metallicity BOSZ template (dashed line without rotational broadening applied and solid line with rotational broadening), with dashed lines showing the original data and solid lines showing net calcium emission with template subtracted. The broad absorption wings of calcium are better represented by the high-metallicity BOSZ template than by the low-metallicity one shown in Figure \ref{pic:metals}.}
	\label{pic:calcium}
\end{figure}

Ca triplet lines (CaII 8498.02, 8542.13 and 8662.14\AA) in the stellar photosphere are significantly broader than other lines, so the wings of the absorption that likely originate in the stellar photosphere are more straightforward to identify. This makes it possible to look closer into the kinematic structure of the emission. The photospheric absorption is noticeably deeper than what is suggested by the low-metallicity template (Figure \ref{pic:metals}), therefore we try the high-metallicity BOSZ template which represents absorption wings much better, as shown in Figure \ref{pic:calcium}. Applying a $\sigma_v=15$ km s$^{-1}$ rotational broadening to the template and subtracting it, we find that calcium emission potentially has a much broader kinematic structure than just the obvious emission line peak. The exact shape can range from double-peaked (CaII 8498.02\AA) to red-winged (CaII 8662.14\AA). The latter profile is narrower than but reminiscent of the kinematic structure of Na I D, further reinforcing the idea that metal emission and Na D absorption originate in the same material. The template-subtracted emission is sensitive to the exact template chosen, so the double-peaked Ca finding remains speculative.

\subsection{Extinction curve}

As seen in the AAVSO data in Figure \ref{pic:LC}, in the dimmest part of the lightcurve the $I$-band points are above the $V$-band points, but during the egress all points come close to being on top of each other, suggesting that the object was redder during the dimmest parts of the eclipse than during the brighter parts of the egress. We bin the lightcurve into temporal bins of variable duration depending on the rate of the evolution (smaller bins for egress and longer bins for the dimmer part of the curve), median-average the photometry within these bins and find that $V-I$ ranges between $0.75-0.85$ mag at $V\simeq 17$ mag (eclipse) and $0.45-0.55$ mag at $V\simeq 14$ mag (egress). Qualitatively this trend is consistent with dust extinction which is stronger at shorter wavelengths that at longer wavelengths, as in typical interstellar medium dust. 

\begin{figure}
	\centering
	\includegraphics[width=\linewidth]{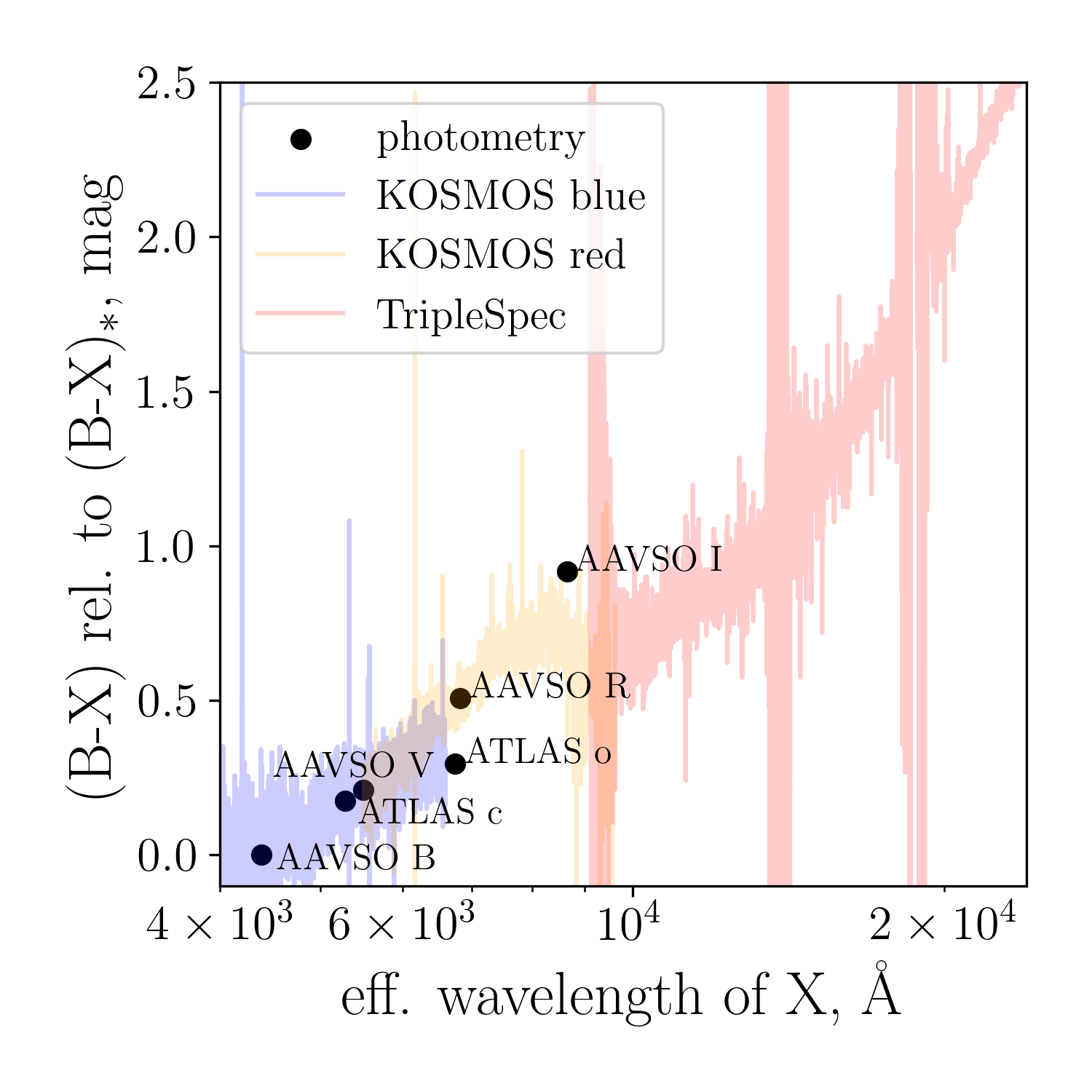}
	\caption{Reddening (relative to the synthetic stellar colors pre-occultation calculated using high-metallicity BOSZ template) as a function of wavelength; the vertical axis is identical to $A_B-A(\lambda)$. It is calculated based on the median photometry between MJD=60650 and 60750, during the dimmest stretch of the event, and based on epochs 2 and 3 of APO spectroscopy. Spectroscopic data for KOSMOS red and Triplespec are arbitrarily offset in the vertical direction to stitch with the shorter wavelength because absolute photometry is unavailable. KOSMOS blue is anchored by the value in the $B$ band (i.e., $A_B-A(\lambda)$ directly measured from the spectrum is shown). }
	\label{pic:extinction}
\end{figure}

To evaluate the extinction curve quantitatively, we median-average all AAVSO and ATLAS photometric datapoints during the dim part of the lightcurve. Multi-color observations with the same set of filters are not available outside of the event, therefore we compute the synthetic photometry from the BOSZ templates (the conclusions do not meaningfully change depending on whether we use the low-metallicity or the high-metallicity template) and the filter curves tabulated in \citet{tonr18} for ATLAS orange and cyan bands, and by the Spanish Virtual Observatory \citep{rodr12, rodr20, rodr24} for the AAVSO Johnson bands. Comparison of the observed colors during the dimmest part of the eclipse and the synthetic colors expected for \jj\ outside of the eclipse reveals the extinction curve shown in Figure \ref{pic:extinction}. 

On the 2nd and 3rd nights of APO observations, we attempted to spectrophotometrically calibrate our data by observing a nearby standard star close in time to scientific observations and by placing the slit at the parallactic angle to minimize slit losses due to atmospheric refraction. The conditions were not photometric as the seeing was quite variable on both occasions. For example, for the second APO epoch the nominal flux through the slit for the standard star observation varied by 65\% even over the span of short exposures, but the relative spectrophotometry of that same star was in agreement to a few per cent. As both epochs were within the deepest parts of the eclipse, we extract the extinction curve directly from our spectroscopic observations for comparison with photometry. For KOSMOS data, we calculate 
\begin{equation}
 A(\lambda)=-2.5\times \log_{10}\left(\frac{F_{\lambda}({\rm obs})S_{\lambda}({\rm model})}{F_{\lambda}({\rm model})S_{\lambda}({\rm obs})}\right).    
\end{equation}
Here $F_{\lambda}({\rm obs})$ is the observed spectrum of the target, $F_{\lambda}({\rm model})$ is its pre-occultation BOSZ spectrum, and $S_{\lambda}$ are the corresponding values for the nearby standard star. This expression can be applied to both flux calibrated and flux uncalibrated datasets and with and without telluric corrections, because for observations with the same setup and same airmass, the ratio $F_{\lambda}({\rm obs})/S_{\lambda}({\rm obs})$ takes care of the wavelength-dependent flux sensitivity of the detector and of the telluric absorption. For TripleSpec data where both flux and telluric calibrations are applied during the reduction process, we calculate $A(\lambda)=-2.5\times \log_{10}\left(F_{\lambda}({\rm obs})/F_{\lambda}({\rm model})\right)$. The model spectrum consists of the BOSZ template which is augmented by a smooth function of wavelength by 25\% to account for the rise of the pre-occultation SED between $H$ and $K_S$ bands (Figure \ref{pic:sed}). The results are shown in Figure \ref{pic:extinction} for a qualitative comparison and are in good agreement with photometry. 

We can use the measured extinction curve to calculate the nominal ratio of the extinction to reddening, $R_V\equiv A_V/E(B-V)$, with $A_V\simeq 4$ mag and $E(B-V)=0.30$ mag as directly measured by the AAVSO photometry. The value $R_V\simeq 13$ we measure for the extinction in the \jj\ occulter is significantly higher than the canonical interstellar medium value of $R_V=3.1$; this conclusion is insensitive to which template is used. This could imply that the occulter is lacking small ($\sim 0.1$\micron) grains that produce significantly more extinction in the blue than in the red for a given overall column density of material \citep{wein01}. Indeed, in an extreme example, if the occulter consisted of a pile of macroscopic rocks, it would block all wavelengths equally, producing "grey" (i.e., wavelength independent) extinction with an infinitely high value of $R_V$. Therefore, the measured extinction curve with relatively modest reddening compared to the overall 4-magnitude extinction can be evidence of dust size distribution poor in small grains. Top-heavy particle size distributions are sometimes seen in molecular clouds, young stellar objects or evolved stars \citep{mura10, miot14, chen16, woit19} or alternatively of a top-heavy collisional cascade that is not producing small particles at a sufficient rate \citep{yang24}. Much of the information on the particle size distribution in these sources comes from the modeling of their submm spectral energy distributions where the spectral index depends on the particle sizes \citep{woit19}. If the submm emission from \jj\ is ever detected in the future, it will provide additional constraints on the size distribution. 

Using BHTOM photometry, we can calculate the extinction curve directly from the data, without involving the model photosphere, because there are a handful of data points available at the end of the egress that we take to be the photometry of the star without the occulter. BHTOM $B$-band photometry is only available after MJD=60765, when the lightcurve is already on the rise toward the egress. We fit piece-wise polynomials to the late-time $B-$ and $V-$band lightcurves to minimize the effects of photometric uncertainties. We obtain the total change in color $\Delta(B-V)=-0.18$ mag (object becomes bluer during the egress) and the total change in brightness $\Delta V=-3.57$ mag (object becomes brighter during the egress), with $R_V\simeq 20$. There is scatter around the best-fit models at the level of 0.1 mag, and therefore with such a small change of color it is likely that the uncertainty on the $R_V$ measurement is quite large, $\pm 10$. But overall the large $R_V$ value from BHTOM is qualitatively consistent with that derived above and is consistent with the lack of small grains. Based on their own photometric follow-up during the occultation, \citet{fore25} reach similar conclusions regarding the dust opacity of the occulter -- that it is nearly wavelength independent -- and present specific dust models compatible with these observations. 

\section{Physical model of the system}
\label{sec:model}

\subsection{The age of the system}

Pre-occultation observations from {\it Gaia} and other photometric optical surveys indicate that the star is a $>2$ Gyr-old main-sequence star with no clear signs of youth in its optical photometry or Galactic kinematics, and it is not in an open cluster. Rotational broadening measured during the occultation from the transmitted stellar spectrum is consistent with the age and the mass of the star derived from photometry. The stringent limit on photospheric lithium absorption further suggests that the star is not young.

However, infrared observations indicate that there is a luminous dust disk -- normally associated with young stars -- that extends from the dust sublimation radius out to at least 1.4 AU from the star. Furthermore, \jj\ displays broad H$\alpha$ and Pa $\beta$ emission which likely originates close to the star. Such emission is also commonly seen in young stars. Thus there is tension between different available age indicators. 

The infrared excess with $L_{\rm IR}/L_{\rm bol}=9-14\%$ places \jj\ on the upper end of the distribution for the interesting and possibly heterogeneous sample of dusty main-sequence stars (\citealt{sank21} and references therein). An interesting possibility that might simultaneously explain multiple features of \jj\ is a collision of planets or substellar companions. Such events can result from dynamical instabilities in planetary systems even long after the nominal planet formation stage \citep{juri08, frel19}, including at $\gtrsim 1$ Gyr timescales \citep{schm25}. A collision  would populate the system with gas and dust and help explain the difference between the photospheric and kinematic age diagnostics and the presence of infrared excess and hydrogen emission. Another possibility is the planet-planet scattering event, which can result in a cloud of debris if the planets were accompanied by satellite systems due to the collisions and tidal disruptions of the satellites \citep{must24}, although it may be more challenging to explain the origin of the observed hydrogen in this scenario if the satellites are gas-poor. We further discuss below how the collisional scenarios can help explain some of the features of \jj.  

\subsection{Orbital configuration}

We primarily explore a model in which the star is occulted by a transiting circumsecondary disk which is gravitationally bound to an unseen object and which in turn is conducting orbital motion around the star. We adopt the orbital period $P_{\rm orb}=16,000$ days and the duration of the eclipse $\tau_{\rm ecl}=254$ days between mid-September 2024 and end of May 2025, as well as the stellar parameters listed above -- most importantly, $M_*=1.4 M_{\odot}$. If the orbit is circular, then the occulter is at $a_{\rm orb}=13.9$ AU from the star and moving with orbital velocity $v_{\rm orb}=9.5$ km s$^{-1}$. 

\begin{figure}
	\centering
	\includegraphics[width=\linewidth]{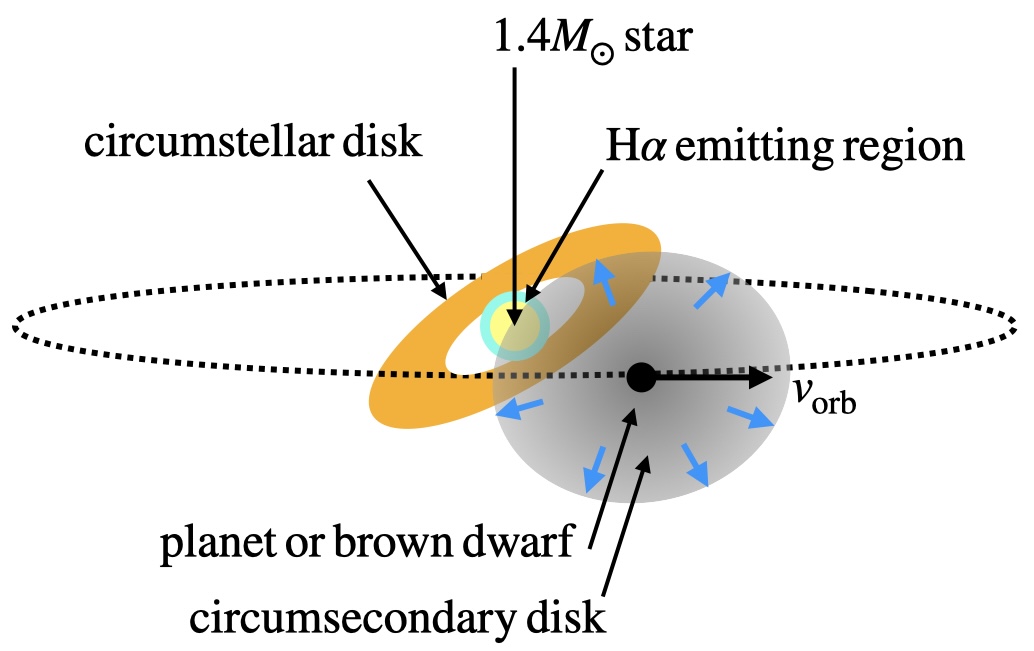}
	\caption{A schematic representation of our model for \jj\ (not to scale). The star (yellow) and the broad H$\alpha$ emission (cyan) are the most compact sources of emission. The infrared-emitting circumstellar disk (orange) extends from sublimation distance ($\sim 0.15$ AU) out to distances that are poorly constrained by observations, $1.4-19$ AU depending on the model assumptions, and fits suggest that it is not aligned with the 14 AU orbit of the occulter (dotted line) which is close to edge-on. A planet or brown dwarf secondary is surrounded by a gaseous, puffy disk (grey) which produces Na D absorption and multiple metal emission lines that show kinematics inconsistent with passive orbital motion, so this gas may be in an outflow (blue arrows).}
	\label{pic:cartoon}
\end{figure}

If the occulter had a sharp edge within which the material immediately has a high optical depth, then the time for ingress and egress should be just the time it takes for the edge to cover the stellar diameter at the orbital velocity, $2R_*/v_{\rm orb}$, which is about 3 days. Both the ingress and the egress happened on significantly longer timescales, 15-20 days. It follows that the ingress and the egress timescales are determined not by the stellar radius occulted by a sharp edge but by the gradient of the column density in the occulter. Adopting $\tau_{\rm ingress}\sim 20$ days, we find that $2\tau_{\rm ingress}/\tau_{\rm ecl}=16$\% is the `fuzziness' of the occulter's edge, i.e., the fraction of its size over which its column density increases from 0 to its near-maximal value. The duration of the ingress and egress can also be altered by eccentricity; $e>0.9$ would be required to slow the ingress by a factor of 5.

If the occulter's orbit is exactly edge-on and if the occulter is circular in projection on the plane of the sky, then during the occultation we are seeing the occulter's maximal extent, and the eclipse duration is directly related to the physical size of the circumsecondary disk, $\tau_{\rm ecl}v_{\rm orb}=2R_{\rm CSD}$, so $R_{\rm CSD}=0.7$ AU. If the disk radius corresponds to a fraction $\xi$ of the Hill radius, then we can find the mass of the planet:  
\begin{equation}
    m_{\rm secondary}=3M_*\left(\frac{R_H}{a_{\rm orb}}\right)^3=
    3M_*\left(\frac{\pi\tau_{\rm ecl} }{\xi P_{\rm orb} }\right)^3.
    \label{eq:hill}
\end{equation}
Theoretical calculations and numerical simulations suggest that the circumplanetary disk can be truncated anywhere between $0.3-0.4R_H$ \citep{ayli09, mart11} and $1R_H$ \citep{szul16, mart23}, with latter values requiring significant gas pressure support. The physical size of the occulter is set by the duration of the transit. If it fills the full Hill sphere ($\xi=1$), then the minimum required mass of the gravitating object in \jj\ is $0.5 M_{\rm Jup}$, in the planetary regime. If instead the disk is truncated at $0.4R_H$ or $0.3R_H$, then the actual Hill radius is larger than the duration of the occultation suggests, and the required minimum secondary mass is larger -- 8$M_{\rm Jup}$ or 19$M_{\rm Jup}$, correspondingly. The Hill-based range of $0.5-19M_{\rm Jup}$ is the lower limit on the mass of the companion, necessary to gravitationally contain the disk if the companion is on a circular orbit. If the orbit is eccentric, the mass of the companion would need to be larger to retain the disk at the pericenter, plus the disk's size $\tau_{\rm ecl}v_{\rm orb}$ could be underestimated using circular orbital velocity implicit in eq. (\ref{eq:hill}). Depending on the thickness of the disk, on its orientation relative to the orbital angular momentum, and on the impact parameter (the projected distance between the star and the disk center), we might not be seeing its full extent during the occultation and therefore the observed $\tau_{\rm ecl}$ may not represent its full size. 

No binarity flags are raised either by the default {\it Gaia} radial velocity analysis pipeline or by the astrometric analysis pipeline. An independent re-analysis of the pre-occultation {\it Gaia} excess radial velocity noise by \citet{chan22} indicates that \jj\ has only a 7\% probability of being a single star. Our own observations with GHOST during the occultation indicate a radial velocity of 32.9$-$36.7 km s$^{-1}$ based on the fits to the centroids of the pure photospheric absorption lines (depending on which set of lines is chosen), consistent with the {\it Gaia} value within its stated error. The lack of detection of radial velocity variations between {\it Gaia} and GHOST -- $<$4 km s$^{-1}$ in 5 years at 13.9 AU from the star is not as constraining as photometric limits. \citet{fore25} estimate the mass of $0.25M_{\odot}$ if all of the near-infrared excess during the occultation is due to the companion, which can be treated as an upper limit on the companion mass if there are other contributions to the near-infrared emission or if the companion is a result of a recent collision of substellar objects and is therefore young and still cooling. An unresolved M-dwarf secondary that is consistent with the optical data, parallax, and extinction as well as physically allowed by stellar evolution would contribute up to 3\% of the pre-occultation flux at 1 micron and is not ruled out. There is a faint source 3\arcsec\ away from \jj\ seen in our acquisition images which is present in {\it Gaia} catalog, but as there are no kinematic measurements, it is unknown whether it is physically related to our target. If it is a wide-binary companion, the projected distance is about 3,000 AU.

Similar mass calculations for secondaries were performed by \citet{scot14} for another periodic dimming transient, also analyzed by \citet{dong14}. In this object, $\tau_{\rm ecl}=16$ days, $P_{\rm orb}=468$ days and $M_*=4M_{\odot}$. For the same assumptions on the truncation radius $\xi$, the secondary is about 30 times more massive than that in \jj, i.e. a brown dwarf or a low-mass star. The lightcurve within the eclipse is significantly structured, with an initial flat plateau and a deeper short dip toward the end of the eclipse. An eerily similar lightcurve is seen in another disk eclipse candidate by \citet{ratt15}, although in this case the eclipse lightcurve is known to change noticeably from one cycle to the next. The eclipse duration is $\tau_{\rm ecl}=100$ days and the period is $P_{\rm orb}=1277$ days. The star is a K giant with an uncertain mass; taking $M_*=1M_{\odot}$ we find that the secondary's mass is about 80 times that of the secondary in \jj, likely a low-mass star. Therefore, among occultations with known periods and assuming that this subclass of occultations is by Hill-limited disks around secondaries described by eq. (\ref{eq:hill}), \jj\ stands out as the lowest mass disk-confining secondary, potentially in the planetary mass range, and as the one with the most symmetric and smooth lightcurve. Unfortunately for some one-off events like those of \citet{mama12} an \citet{rapp19} the period is unknown and therefore the mass of the secondary cannot be calculated using eq. (\ref{eq:hill}). 

Before the occultation, the star had a significant infrared excess which we fit with the optically thick disk model of \citet{jura03}. This model is idealized; in particular, it assumes that the disk is not warped and therefore no direct emission illuminates it at large distances from the star. Taking $T_{\rm out}=100$ K as the coldest reliably detected thermal dust component, we use the \citet{chia97} annular radius-to-temperature mapping to find that this emission originates at $a_{\rm out}=1.4$ AU from the star. This is significantly more compact than what would be obtained from considering a dust grain's thermal equilibrium with the flux it receives from the star, $T\sim 100$ K at $a_{\rm out}\sim 19$ AU. This order-of-magnitude difference in the estimates of the size of the mid-infrared emission region is due to optical depth effects. If the disk is perfectly planar and shields its outer parts from stellar radiation, then the dust can maintain cool temperatures close to the star. But if there are any warps in the disk, then cooler temperatures occur much further out. This dramatic difference in the estimates of the emission size underscores the difficulty of determining the extent of the circumstellar disk relative to the size of the occulter or even relative to the occulter's orbit. 

The inner size of the circumstellar disk can be estimated from the \citet{chia97} model and from the thermal equilibrium at the dust sublimation temperature ($\sim 1200$ K). These values are in the range of $0.06-0.16$ AU and are much smaller than the size of the occulter. Therefore, the near-infrared part of the SED of \jj\ should also be occulted with an occultation duration similar to that in the optical, whereas the mid-infrared emission may remain at least in part unaffected by occultation as it is produced on scales larger than the occulter. Unfortunately no MIR photometry are available during the occultation. The extinction curve we estimate from TripleSpec data in Figure \ref{pic:extinction} already takes into account that the host dust seen in the $K_S$ band is likely entirely behind the occulter during the observation, as this component is incorporated into the model spectrum used for the extinction curve calculation. 

The projection factor $\cos i$ derived for the infrared-emitting circumstellar disk is not particularly small, so there is no evidence that the circumstellar disk is close to edge-on -- whereas the occulter's orbit is edge-on or close to it. If we take the results of the disk SED fitting at face value, then we deduce that the inner disk is misaligned with the orbit of the outer occulter. Such misalignments can occur through sweeping secular resonances during disc dispersal \citep{mats17, owen17}. An example of an outer planet misaligned with the inner disk is presented by \citet{nguy21}, although the system architecture is quite different in their case: the star is a close binary, the planet is much further away, and the system is young. 

\subsection{Conditions within the occulting structure}

The Na I D absorption seen during the occultation -- in excess of its strength in the stellar photosphere and of interstellar contribution -- together with the gaseous metal emission lines suggest that the occulter is a gas-rich structure. The smoothness of the lightcurve during the event is another piece of evidence for the occulter being gas-rich. A contrasting example is that of \citet{mama12} -- the lightcurve of their event shows extreme variations in brightness on short timescales, suggesting that the occulting disk is razor-thin, which is inconsistent with expectations for gas-rich circumstellar disks \citep{ayli09}. The greater the aspect ratio (i.e., the more the occulter tends toward spherical symmetry), the more difficult it is to produce features or asymmetries in the lightcurve \citep{scot14}. The lightcurve of \jj\ is significantly more smooth and symmetric than those seen by \citet{scot14, ratt15} and \citet{rapp19}, with only 0.1 mag deviations relative to a symmetric fit to the eclipse and no evidence for statistically significant stochastic variability relative to this function, nor any evidence for gaps or rings which would manifest as correlated signals before and after the transit mid-point. It is possible to produce a symmetric lightcurve by a geometrically thin disk if its center transits with a very small impact parameter relative to the star, but a more natural and less fine-tuned explanation is that the structure is geometrically thick. All of these arguments favor a gas-rich nature for the occulter in \jj, though the small (0.1 mag) variations relative to the perfect lightcurve symmetry rule out an occulter that's completely spherically symmetric, as well as a disk-like occulter with a 0 impact parameter. 

\jj\ presents many interesting puzzles compared to numerical simulations of circumplanetary disks. The flatness of the light profile during the eclipse is inconsistent with models where the surface density steeply declines with the distance from the planet (or the brown dwarf), and we did not detect any sign of central rebrightening which would have indicated depletion of disk material onto the object. However, this signature can be easily missed depending on the disk's thickness, inclination and impact parameter. 

We use our Na D observations to estimate the amount of gas in the occulter. For low optical depth, the equivalent width of the Na I D absorption lines should be proportional to the extinction, with the coefficient of proportionality dependent on the ratio of dust to gas-phase sodium. This proportionality breaks down for equivalent widths well under those seen in \jj\ \citep{pozn12}. The observed equivalent widths of the D$_2$ and D$_1$ lines are, correspondingly, 0.7\AA\ and 0.4\AA, after correcting for the photospheric component of Na I D absorption. The empirical relationships for the interstellar medium (ISM) from \citet{pozn12} suggest corresponding $A_V$ values anywhere between 0.5 and 13 magnitudes, a wide range that takes into account the uncertainty in our Na I D measurement and the large systematic uncertainty of the scaling relationships in the non-linear regime (and the validity of these relationships at high extinction values are additionally questioned in re-analysis of \citealt{murg15}). This range does comfortably encompass the observed value $A_V\simeq 4$ mag, and therefore there is no evidence in the data to suggest that the occulter has an amount of gas-phase sodium which is anomalous for the amount of dust in the occulter. 

If we were to use the standard sodium-to-hydrogen ratio and extrapolate the relationships by \citet{murg15} derived for low optical depth ISM well beyond their regimes of validity, then we would estimate the hydrogen column density of $N_H=1.5\times 10^{21}$ cm$^{-2}$. This is the most uncertain factor in the mass estimate of the gas in the disk because the directly measured column density of gas from the strength of Na I D absorption is only $N_{\rm NaI}=6.4\times 10^{12}$ cm$^{-2}$ \citep{drai11}, so the unknown abundance of metals and hydrogen fraction in the occulter introduce the uncertainty of several orders of magnitude. We next assume that the circumsecondary disk is not a thin edge-on structure and therefore its apparent size (i.e., radius $R_{\rm CSD}$) can yield its area projected on the plane of the sky via $\pi R_{\rm CSD}^2$ to within factors of order unity. Then we would calculate the mass of the neutral hydrogen in the circumsecondary disk to be $m_{\rm proton} N_H \pi R_{\rm CSD}^2=8\times 10^{23}$ g, or 1.1\% of the mass of the Moon. There is no guarantee that the sodium-to-hydrogen ratio in the occulter is anything like that in the interstellar medium -- for example, if the gas is of secondary origin, i.e., due to the collisions of dust particles, then we might expect that the disk is depleted in light elements and that its gas-to-dust ratio is significantly lower than that in the ISM. In that case our estimate serves as an upper limit on the gas mas. 


\subsection{Kinematics of the gas within the occulter and the origin of the system}

The strongest components of Na I D absorption -- blueshifted at 27 km s$^{-1}$ with respect to the star -- have a small velocity dispersion ($\sigma_v=4.5-13$ km s$^{-1}$), and the same kinematics is shared by two dozen metal emission lines, which suggests that they too originate in the occulter. Furthermore, subtraction of the stellar photosphere contribution for calcium triplet, where it can be more confidently identified than in most other lines, reveals that both sodium absorption and calcium emission may share broad components that are at the same systemic velocity as the star or even redshifted relative to it. 

While the similarity of the kinematics of Na D and metal lines suggests origin in the same material, this similarity is actually surprising, because their spatial distribution probed by the GHOST spectrum can be quite different. To make faint emission lines from some component of the system visible, all that is needed is to block the star to provide a better contrast, and otherwise the emitting region does not have to be aligned with the star. Therefore, metal emission lines may in principle originate anywhere or everywhere in the occulter or even in other parts of the circumstellar disk. In contrast, the Na I D absorption probes the kinematics of the gas only in one spot -- exactly along the line of sight to the star. 

First we consider the possibility that the Na D and metal emission at $-27$ km s$^{-1}$ with respect to the star are co-moving with the occulter and reflect the occulter's radial velocity. For a circular orbit, the radial velocity during an occultation is always zero. Therefore, the only way to generate a 27 km s$^{-1}$ blueshift relative the star using the orbital motion at the semi-major axis of 14 AU (fixed by the historic periodicity) would be to have eccentricity $>0.95$ and to have the observer fortuitously aligned not far from the pericenter where an eccentric orbiter spends very little time. It seems impossible to gravitationally contain such a highly eccentric tail of debris, and it is further implausible that the lightcurve would be so smooth and symmetric. The double-peaked emission profile of photosphere-subtracted CaII 8498.02\AA\ (if it is real and not a result of a poor template match) might suggest that the occulter has a radial velocity of $-13$ km s$^{-1}$ with respect to the star and it generates two peaks of emission at $\pm 13$ km s$^{-1}$ with respect to its own frame (either due to rotation or winds). But even a 13 km s$^{-1}$ radial velocity offset between the occulter and the star requires eccentricity $>0.81$ and a fortuitous alignment of the observer with respect to the pericenter. If the disk was rotating with a characteristic velocity 13$-$27 km s$^{-1}$ at the distance of 0.1 AU from the gravitating object to explain MagE and GHOST data, then the mass of the secondary would need to be 19$-$76 M$_{\rm Jup}$. 

Another possibility is that the secondary object has launched an outflow with a characteristic velocity 13$-$27 km s$^{-1}$. Then the redshifted part of the outflow is preferentially more extincted by the occulter itself and is therefore appreciably fainter, although may still be visible as the double-peaked profile of calcium. While this velocity scale is not out of question for an outflow from a massive forming planet \citep{quil98}, it is still challenging to explain the physical state of the gas -- the presence of metal emission lines with characteristic excitation temperatures of tens of thousands of degrees -- and especially the spatially distributed Na I D absorption. Inflows and outflows of ionized gas are expected near the planet via magnetospheric, photoevaporative or magneto-centrifugally driven mechanisms \citep{gres13, zhu15, hart16}, and the magnetospheric accretion can produce a variety of kinematic signatures including blueshift \citep{hart16}. But the outflows would need to affect a spatially extended region within the circumsecondary disk in order to explain the kinematic substructure of Na D. So the similarity of the kinematics of Na I D and the of the metal lines is quite puzzling in any scenarios that can plausibly explain metal lines by processes near the planet. In the one known case (PDS 70; \citealt{haff19}) of two forming protoplanets directly imaged in the outer reaches of their exoplanetary system, their H$\alpha$ emission is redshifted relative to the star by $\sim 30$ km s$^{-1}$, well in excess the orbital velocity scale. This observation directly confirms that velocities of a few tens of km s$^{-1}$ and excitation temperatures similar to those seen in our source can be generated during planet formation, but does not shed light on the possible spatial extent of this emission or on the origin of blueshift in \jj. 

\citet{fore25} report that \jj\ was strongly ($P\sim 4\%$) polarized during the occultation in the optical range. If the mechanism of the polarization is the same as in the interstellar medium of the Galaxy -- dichroic extinction \citep{davi51} -- then the grains must be aligned, likely magnetically. The degree of polarization in the ISM is $P_{\rm ISM}\le3\% \times A_V$ \citep{serk75}, so if the dust in \jj\ was ISM-like, then values of $P<12\%$ would not be out of reach for the dichroic extinction mechanism. Therefore, the blueshifted emission lines presented here and the polarization discovered by \citet{fore25} potentially suggest that the occulter is experiencing a magnetized wind. Unfortunately, this is not a unique explanation because similar levels of polarization can be achieved without magnetic field by multiple scatterings in the disk if its geometry is favorable \citep{bast88}.

It would be interesting to know whether there were any variations in the kinematics of the Na D during the occultation. For example, if the kinematics of Na D are determined by the wind in the occulter then in the beginning of the occultation the line-of-sight gas velocity has no blueshift as the gas moves close to the plane of the sky, then it shows maximum blueshift during the peak occultation, and then the relative velocity declines again. If the kinematics are due to a rotating disk, then we might expect to see a net change in the dominant Na D velocity relative to the star during the occultation. If the disk angular momentum is aligned with the orbital angular momentum, then we expect to see redshift in the first half of the occultation followed by blueshift in the second half, and a blueshift-to-redshift transition if the angular momenta are anti-aligned. The two highest quality spectra, MagE and GHOST from epochs 4 and 5, both taken in the second half of the occultation and relatively close to the mid-point, show blue-shifted Na D with no net kinematic differences between each other. But the centroids of Na D from KOSMOS data obtained during the first half of the occultation are unfortunately ambiguous at the 20 km s$^{-1}$ level, so while there is no evidence for a change from a redshift to a blueshift during the occultation, we cannot rule out this possibility. 

Colliding or scattering planets or sub-stellar companions would produce a disk tracing the angular momentum in the center of mass of the impact, which would naturally explain a wide range of disk inclinations relative to the orbital angular momentum around the star. A planetary collision in \jj\ would populate the entire system with debris that would explain its significant infrared excess, and the occulter could represent the main cloud of debris. Given that the star shows hydrogen emission and the circumsecondary disk shows metal emission, at least one of the objects involved must have had a rocky core and at least one object must have had a gaseous atmosphere. The ratio of H$\alpha$ equivalent width to the relative circumstellar disk strength $L_{\rm disk}/L_*$ can be taken as a very rough proxy for whether the circumstellar disk is hydrogen-rich or dust-rich. If \jj\ is indeed a result of a collision, then it is a unique system without a good available comparison system. It is then interesting that \jj\ is in the lowest quartile for this ratio (i.e., relatively hydrogen-poor) compared to the young stellar objects in \citet{cabr90}, but within their range, so in the planet collision scenario the composition of the debris ended up not far off that of young protoplanetary disks.

It seems likely that there must be a surviving object gravitationally containing most of this debris to produce repeated eclipses of similar duration; simulations show that a remnant planet or a close binary are likely to form instead of complete destruction \citep{hwan18_planets}. Further dynamical modeling may shed light on whether a cloud of debris can survive for extended periods without gravitational confinement and whether planet collisions are likely to produce inner circumstellar disk misaligned with the orbit of the main cloud of the debris. In the main occulting cloud, the debris may be collisionally grinding down, producing gas emission \citep{roge25} which would be in a large-scale (spanning the entire occulter), pressure-driven outflow. This scenario may be able to explain the kinematics of the gas and may be qualitatively consistent with the unusual extinction curve which suggests paucity of small grains (grain growth within the disk is also possible). Dynamical instabilities in planetary systems have been used to explain infrared brightening events and optical occultations in a number of sources \citep{meng14, kenw23, mars23}. The difference between these events and \jj\ was that the collision in the \jj\ system happened before 1937 and produced a lot more dust, likely because the instability involved more massive objects.

\subsection{Alternative models}

In addition to the circumsecondary disk scenario discussed here, \citet{fore25} present two alternative scenarios: a circumstellar disk precessing due to a $\le 0.25M_{\odot}$ companion located fully outside of the disk, and a warped or precessing circumbinary disk. For example, a large warped disk of CQ Tau, which may be a representative of a particular class of young stellar objects, UX Ori, is directly confirmed with CO kinematics \citep{wolf21}. There are suspicions that a planet must be responsible for creating and maintaining the disk substructures responsible for occultations in CQ Tau, but no specific companion has yet been detected \citep{hamm22}. Both \jj\ and CQ Tau have IR excess and dim by several magnitudes on similar timescales (although the periodicity of CQ Tau inferred from the optical lightcurve is ambiguous, 21 or 10 years; \citealt{shak05, grin23}). There are also glaring differences, in particular the lightcurve of CQ Tau is extremely irregular by comparison to the lightcurve of \jj. On the basis of the smoothness and the symmetry of the occultation, the lack of variability outside the eclipse, and the regularity of the three known eclipses (to within the limits of the historical data), we focus on the circumsecondary disk scenario.

Regardless of the exact geometric configuration in \jj, there is tension between the presence of gas and dust, which suggest youth, and all other age diagnostics, which suggest that the star is $>2$ Gyr old. If the star is an old main-sequence one, then regardless of whether \jj\ is occulted by a circumsecondary disk, circumstellar disk, or circumbinary disk, the origin of the gas and dust in this system likely requires a rare event like a collision of sub-stellar companions or a disruption of a planet.

\section{Conclusions}
\label{sec:conclusions}

In this paper we present results from multi-epoch spectroscopy of ASASSN-24fw or \jj, a deep dimming event of a star with mass 1.4$M_{\odot}$. Historic data in combination with the high-quality photometric follow up of the 2024-25 event demonstrate that the dimming events are periodic with a period of 44 years and that they last $\sim 9$ months. Assuming that the dimming event is due to a tidally-truncated disk around a secondary object in the orbit around the star \citep{mama12, dong14, scot14, rapp19}, we estimate its orbital separation of 14 AU, the radius of the occulter of 0.7 AU, and the mass of the object gravitationally confining the occulter to be between a few Jupiter masses and $0.25M_{\odot}$, the upper limit by \citet{fore25} from SED fitting during the dimming.

Using optical spectroscopic observations, we discover deep Na I D absorption, well in excess of the expectations from a normal stellar photosphere, H$\alpha$ emission, and two dozen metal emission lines, all with complex kinematics exquisitely measured by a high-resolution Gemini GHOST spectrum. The origin of the kinematics of Na I D absorption, H$\alpha$ emission and metal emission lines is not completely clear. Our best guess, based on the width of H$\alpha$ (hundreds of km s$^{-1}$), is that it originates within the inner disk close to the star and not in the occulter. In contrast, the depth of Na I D absorption is consistent with the optical extinction during the event and is therefore likely associated with the occulter. The kinematic similarity between Na I D and metal emission suggests that they, too, originate in the occulter. The velocity scale of this emission, which has a component blueshifted by 27 km s$^{-1}$ with respect to the star, is not easily explained by the occulter's orbital motion around the star. An intriguing possibilities are winds from the circumsecondary disk or disk's rotation, but understanding whether the gas physical conditions or the winds' spatial and velocity scales are plausible will require further theoretical modeling. 

The extinction curve measured during the event suggests that the grain size within the occulter is larger than that in the interstellar medium. The object's Galactic kinematics and optical spectrum before occultation both suggest a main-sequence star in intermediate stages of its main-sequence evolution, but the source shows strong and somewhat time-varying infrared excess which would normally be characteristic of disks around young stars. These observations might be reconciled in the scenario where the \jj\ system suffered a dynamical instability and a planetary or substellar collision that populated the entire system with dust and gas and produced the main cloud of debris we see as the occulter with top-heavy size distribution unlike that in the interstellar medium. 

Some of the remaining questions should be clarified by a high-quality, high-resolution post-occultation spectrum which would unambiguously establish which features originate near the star and which originate in the occulter. There are unfortunately no opportunities to go back to the first half of the occultation and to obtain a better quality spectrum to measure the kinematic changes during the occultation. However, there are some interesting options for future follow-up observations that may still shed light on the architecture of the system, on the mass of the secondary, and on the dynamics of the circumsecondary disk. Submm observations of molecular gas, as well as of atomic and ionic emission lines in the submm range which is unaffected by the contamination by the stellar photosphere, could be useful in mapping the large-scale circumstellar disk and in measuring the kinematics of the the circumsecondary disk, and submm observations of dust continuum can provide additional insights into dust size distribution and therefore on the origin of the disk. 

\begin{acknowledgments}

N.L.Z. is thankful to J.Budaj, C.Espaillat, L.Hillenbrand, J.Munday, A.Rest, B.Shappee, and K.Shukawa for useful and timely information and to S.Gezari and C.Ingebretsen for generous offers of observing time and observing support. G.A.P. is supported in part by the JHU President's Frontier Award to N.L.Z.. The research of J.M. was supported by the Czech Science Foundation (GACR) project no. 24-10608O and by the Spanish Ministry of Science and Innovation with the grant no. PID2023-146453NB-100 (PLAtoSOnG). J.E.O. is supported by a Royal Society University Research Fellowship. This project has received funding from the European Research Council (ERC) under the European Union’s Horizon 2020 research and innovation programme (Grant agreement No. 853022). A.G. was partially supported by a program of the Polish Ministry of Science under the title ‘Regional Excellence Initiative’, project no. RID/SP/0050/2024/1. K.R.H. acknowledges the Smithsonian Institution for funding her as a Submillimeter Array Fellow. E.P. and J.Z. acknowledge funding from the Research Council of Lithuania (LMTLT, grant No. S-LL-24-1). {\L}.W. acknowledges Polish NCN DAINA grant No. 2024/52/L/ST9/00210. 

This research is based on observations obtained at the international Gemini Observatory, a program of NSF NOIRLab, which is managed by the Association of Universities for Research in Astronomy (AURA) under a cooperative agreement with the U.S. National Science Foundation on behalf of the Gemini Observatory partnership: the U.S. National Science Foundation (United States), National Research Council (Canada), Agencia Nacional de Investigaci\'{o}n y Desarrollo (Chile), Ministerio de Ciencia, Tecnolog\'{i}a e Innovaci\'{o}n (Argentina), Minist\'{e}rio da Ci\^{e}ncia, Tecnologia, Inova\c{c}\~{o}es e Comunica\c{c}\~{o}es (Brazil), and Korea Astronomy and Space Science Institute (Republic of Korea). Data were processed using DRAGONS (Data Reduction for Astronomy from Gemini Observatory North and South). The authors are particularly grateful for the opportunity to use the Director's Discretionary Time program on a short notice. This is research is based on observations obtained with the Apache Point Observatory 3.5-meter telescope, which is owned and operated by the Astrophysical Research Consortium. The authors are grateful to the observatory leadership for the opportunity to use Open Time on a short notice. This paper includes data gathered with the 6.5 meter Magellan Telescopes located at Las Campanas Observatory, Chile. 

BHTOM.space is based on the open-source TOM Toolkit by LCO and received funding from the European Union's Horizon Europe Research and Innovation programme ACME grant No. 101131928 (2024-2028). The 1.2 m Kryoneri telescope is operated by the Institute for Astronomy, Astrophysics, Space Applications and Remote Sensing of the National Observatory of Athens. Some of the observations have been obtained with the 90 cm Schmidt-Cassegrain Telescope (TSC90) in Piwnice Observatory, Institute of Astronomy of the Nicolaus Copernicus University in Toru\'{n} (Poland).

The national facility capability for SkyMapper has been funded through ARC LIEF grant LE130100104 from the Australian Research Council, awarded to the University of Sydney, the Australian National University, Swinburne University of Technology, the University of Queensland, the University of Western Australia, the University of Melbourne, Curtin University of Technology, Monash University and the Australian Astronomical Observatory.  SkyMapper is owned and operated by The Australian National University's Research School of Astronomy and Astrophysics.  The survey data were processed and provided by the SkyMapper Team at ANU.  The SkyMapper node of the All-Sky Virtual Observatory (ASVO) is hosted at the National Computational Infrastructure (NCI).  Development and support of the SkyMapper node of the ASVO has been funded in part by Astronomy Australia Limited (AAL) and the Australian Government through the Commonwealth's Education Investment Fund (EIF) and National Collaborative Research Infrastructure Strategy (NCRIS), particularly the National eResearch Collaboration Tools and Resources (NeCTAR) and the Australian National Data Service Projects (ANDS).  This work has made use of data from the European Space Agency (ESA) mission Gaia (\url{https://www.cosmos.esa.int/gaia}), processed by the Gaia Data Processing and Analysis Consortium (DPAC, \url{https://www.cosmos.esa.int/web/gaia/dpac/consortium}). Funding for the DPAC has been provided by national institutions, in particular the institutions participating in the Gaia Multilateral Agreement. We acknowledge with thanks the variable star observations from the AAVSO International Database contributed by observers worldwide and used in this research.

This research has made use of the SIMBAD database, operated at CDS, Strasbourg, France \citep{wen00}. This research has made use of the VizieR catalog access tool, CDS, Strasbourg, France. The original description of the VizieR service was published in \citet{och00}. This research has made use of NASA's Astrophysics Data System Bibliographic Services. This research has made use of the SVO Filter Profile Service, funded by MCIN/AEI/10.13039/501100011033/ through grant PID2023-146210NB-I00. 

\end{acknowledgments}

\facilities{Gaia, Skymapper, AAVSO}

\software{\texttt{Astropy} \citep{Astropy2013, Astropy2018, Astropy2022}, 
            \texttt{dustmaps} \citep{gre18},
          \texttt{isochrones} \citep{mor15},
          \texttt{R} \citep{r25},
          \texttt{TOPCAT} \citep{tay05}
          }

\bibliography{stars}{}
\bibliographystyle{aasjournalv7}

\end{document}